\documentclass[12pt]{iopart}
\usepackage{iopams,color,bm}
\usepackage{graphics}
\expandafter\let\csname equation*\endcsname\relax

\expandafter\let\csname endequation*\endcsname\relax
\usepackage{ascmac,amssymb,amsmath,amsthm,color}
\usepackage{mathptmx} 
\usepackage{ytableau}
\usepackage[colorlinks=true, linkcolor=red]{hyperref}
\ytableausetup{boxsize=6pt}

\newtheorem{definition}{Definition}

\newcommand{\Slash}[1]{{\ooalign{\hfil/\hfil\crcr$#1$}}}

\begin{document}

\title[]{On integrability of the Killing equation}
\author{Tsuyoshi Houri}
\eads{houri@phys.sci.kobe-u.ac.jp}
\address{Department of Physics, Kobe University, 1-1 Rokkodai, Nada, Kobe, Hyogo, 657-8501 JAPAN}
\author{Kentaro Tomoda}
\eads{k-tomoda@stu.kobe-u.ac.jp}
\address{Department of Physics, Kobe University, 1-1 Rokkodai, Nada, Kobe, Hyogo, 657-8501 JAPAN}
\author{Yukinori Yasui}
\eads{yukinori.yasui@mpg.setsunan.ac.jp}
\address{Faculty of Science and Engineering, Setsunan University, 17-8 Ikeda-Nakamachi, Neyagawa, Osaka, 572-8508 JAPAN}

\begin{abstract}
Killing tensor fields have been thought of as describing hidden symmetry of space(-time)
since they are in one-to-one correspondence with
polynomial first integrals of geodesic equations.
Since many problems in classical mechanics can be formulated as
geodesic problems in curved spaces and spacetimes,
solving the defining equation for Killing tensor fields
(the Killing equation) is a powerful way to integrate equations of motion.
Thus it has been desirable to formulate the integrability conditions of the Killing equation, which serve to determine
the number of linearly independent solutions and
also to restrict the possible forms of solutions tightly.
In this paper, we show the prologation for the Killing equation
in a manner that uses Young symmetrizers.
Using the prolonged equations, we provide the integrability conditions explicitly.
\end{abstract}

\pacs{}

\begin{flushright}
	KOBE-COSMO-17-05
\end{flushright}

\maketitle

\section{Introduction}\label{sec:introduction}
Many problems in classical mechanics can be formulated as geodesic problems
in curved space(-time)s.
For instance, Euler's equations for a rigid body such as Euler top are described
as the geodesic equations on the Lie groups with the corresponding metrics
(see, e.g. Ref.\ \cite{Craig:2008}).
In general relativity, the motion of a particle in a gravitational field is described as geodesics in a curved spacetime.
Less well-known is that, even in the presence of an external force,
the motion of a particle can be framed as the geodesic problem
in an effective curved space(-time) in the same or higher dimensions
(see Ref.\ \cite{Cariglia:2014review} for a review).

In the Hamiltonian formulation, the geodesic equations in a curved space(-time) $M$
with a metric $g_{ab}$ are given by Hamilton's equations
$\dot{x}^a=\partial H/\partial p_a$, $\dot{p}_a=-\partial H/\partial x^a$
with the Hamiltonian $H=(1/2)g^{ab}p_ap_b$.
According to the Liouville-Arnold theorem,
if there exist first integrals that are in involution as many as the number of dimensions,
Hamilton's equations are integrable in the Liouville sense.
However, it is not always easy to find first integrals for a given Hamiltonian.
This has motivated many authors to develop various methods to
investigate the integrability structure of the geodesic equations.

In this paper we focus on Killing tensor fields (KTs) in a curved space(-time),
which have been thought of as describing hidden symmetry of space(-time)
since they are in one-to-one correspondence with polynomial first integrals of the geodesic equations.
A KT of order $p$, denoted by $K_{a_1\cdots a_p}$,
is a symmetric tensor field $K_{a_1\cdots a_p}=K_{(a_1\cdots a_p)}$ which satisfies the Killing equation
\begin{align}
\nabla_{(b} K_{a_1\cdots a_p)} ~=~ 0 \:,
\label{eq:Killing}
\end{align}
where $\nabla$ is the Levi-Civita connection,
and the round brackets denote symmetrization over the enclosed indices.
Square brackets over indices will be used for antisymmetrization.
A metric is a trivial KT, which is always a solution of the Killing equation.
Hence it has been asked whether the Killing equation
has nontrivial solutions for a given metric.
KTs of order 1 are known as Killing vector fields (KVs),
which have been actively studied as spacetime symmetry (isometry).
KTs of order 2 have also been considerably studied in connection
with separation of variables in Hamilton-Jacobi equations \cite{Stackel:1895,Benenti:1975,Benenti:1980}.
In general relativity, a nontrivial KT of order $2$ was discovered
in the Kerr spacetime \cite{Carter:1968,Walker:1970}. Since then,
nontrivial KTs have been investigated in various black hole spacetimes in four and higher dimensions \cite{Davis:2005,Kubiznak:2007,Hioki:2008,Houri:2010,Chow:2010,Kruglikov:2012,Vollmer:2016,Chow:2016}.
In recent years higher-order KTs have attracted much interest \cite{Gibbons:2011,Galajinsky:2012,Cariglia:2014}.

The main purpose of this paper is to write out the integrability conditions
of the Killing equation,
which has a distinguished role to compute
the number of linearly independent solutions
and also to restrict the possible forms of solutions tightly.
To this end, one can employ the so-called prolongation procedure
(see Refs.\ \cite{Bryant:1991,Eastwood:2008,Dunajski:2010}).
For instance, it is well known that the Killing equation for KVs,
\begin{align}
\nabla_{(a}\xi_{b)}=0 \label{KVequation} \,,
\end{align}
leads to the first-order linear partial differential equations for
$\xi_a$ and $\omega_{ab}$ (see, e.g. Ref.\ \cite{Wald:1984}),
\begin{align}
\nabla_a \xi_b &~=~ \omega_{ab} \:, \label{eq:prolong_1pre0}\\
\nabla_a \omega_{bc} &~=~ R_{cba}\:^d\xi_d \:, \label{eq:prolong_1pre1}
\end{align}
where $\omega_{ab}=\nabla_{[a}\xi_{b]}$
and $R_{abc}\:^d$ is the Riemann curvature tensor.
The procedure used here in deriving Eqs.\ \eqref{eq:prolong_1pre0} and \eqref{eq:prolong_1pre1} from Eq.\ \eqref{KVequation} is known as prolongation. Hereafter, such equations obtained by prolongation are referred to as the {\it prolonged equations}.
	
The prolonged equations \eqref{eq:prolong_1pre0} and \eqref{eq:prolong_1pre1} can be viewed as equations for parallel sections of the vector bundle $E^{(1)}\equiv T^*M\oplus \Lambda^2 T^*M$, where $\xi_a$ and $\omega_{ab}$ are sections of $T^*M$ and $\Lambda^2 T^*M$, respectively. Since it is shown that Killing vector fields are in one-to-one correspondence with parallel sections of $E^{(1)}$, the dimension of the space of KVs is bounded by the rank of $E^{(1)}$, i.e. $n(n +1)/2$ in $n$ dimensions.
Similarly, it is shown that KTs of order $p$ are in one-to-one correspondence
with the parallel sections of a certain vector bundle $E^{(p)}$ \cite{Branson:2007,Michel:2014}.
This leads to the Barbance-Delong-Takeuchi-Thompson (BDTT) formula \cite{Barbance:1973,Delong:1982,Takeuchi:1983,Thompson:1986}
\begin{align}
\dim K^p(M) ~\leq ~ \frac{1}{n}~
\binom{n+p}{p+1}
\binom{n+p-1}{p}
=~ \mathrm{rank}\:E^{(p)}
\:, \label{eq:dim_of_KS}
\end{align}
where $K^p(M)$ denotes the space of KTs of order $p$
in an $n$-dimensional space(-time) $M$.
The equality is attained if and only if $M$ is of constant curvature.

In principle, if the prolonged equations are provided,
it is possible to obtain the integrability conditions.
Indeed, they can be found for order 1 \cite{Veblen:1923} and order 2 \cite{Veblen:1923,Hauser:1974,Hauser:1975i,Hauser:1975ii}.
The integrability conditions for $p\geq 3$ were discussed in \cite{Wolf:1998}.
Similar techniques have been applied to other hidden symmetries \cite{Semmelmann:2002,Batista:2014,Batista:2015a,Houri:2015,Batista:2015b,Houri:2016}.

While such a structure of the prolongation for the Killing equation has been realized,
it is not so easy to write out the integrability conditions for $p\geq 2$.
Even writing the prolonged equations out is rather hard
since the expressions become complicated more and more
as $p$ becomes larger.
Thus the integrability conditions for $p\geq 3$
have never been provided explicitly.
The expressions provided in \cite{Wolf:1998} are still complicated
to elucidate the underlying structure.

To challenge this task, our strategy is to take advantage of the structure of prolongation.
We use Young symmetrizers.
A Young symmetrizer is the operator which acts on a tensor field of order $p$ and
projects it onto an irreducible representation of GL$(n)$.
Since Young symmetrizers have many useful properties,
we can derive the prolonged equations of the Killing equation for a general order
and obtain the integrability conditions explicitly for $p=1$, $2$ and $3$.
To our knowledge, the integrability condition for $p=3$ is new in the sense that
the expressions are written explicitly,
and they will be useful to investigate KTs of order 3 for various metrics.
For $p>3$, although we do not write out the integrability conditions,
we make a conjecture on them from several observations for $p=1$, $2$ and $3$.

This paper is organized as follows: In Section \ref{sec:prolong},
we formulate the prolonged equations for the Killing equation in a manner that uses Young symmetrizers.
At the same time, we will briefly explain the definitions and properties of Young symmetrizers there.
For readers who are not familier with them, refer to 
\ref{app:Young} and references therein.
The integrability conditions of the Killing equation are investigated in Section \ref{sec:int_cond}.
We explicitly provide the integrability conditions for order $p=1$, $2$ and $3$.
We also make a conjecture on the integrability condition for a general order and show a method
for computing the dimension of the space of KTs.
In Section \ref{sec:discussion} we conclude with a brief discussion.
Four appendices are devoted to technical details.
\ref{app:Young} describes the properties and some formulas of Young symmetrizers.
\ref{app:prolong} presents a prolongation procedure for the Killing equation up to order $2$.
A derivation of the integrability condition \eqref{eq:int_pp} have been posted in 
\ref{app:d_int}.
In \ref{app:Yano}, we discuss
the Killing-Yano equation by using our analysis.

\section{Prolongation of Killing equation}
\label{sec:prolong}
We shall formulate the prolonged equations for the Killing equation of order $p$
in a manner that uses Young symmetrizers.
For this purpose, we commence by giving a brief definition of Young symmetrizers (see 
\ref{app:Young} for details).

A Young symmetrizer $Y_\Theta$ is the projection operator corresponding to a Young tableau $\Theta$, which makes row-by-row symmetrization and
column-by-column antisymmetrization sequentially. Specifically, it reads
\begin{align}
Y_\Theta
~\equiv~ \alpha_\theta \prod_{C_j \in \mathrm{col(\Theta)}} \hat{A}_{C_j}
\prod_{R_i\in \mathrm{row(\Theta)}} \hat{S}_{R_i}\:,
\label{def:Young}
\end{align}
where $\theta$ is the shape of $\Theta$, i.e. the Young diagram eliminated the numbers from the tableau $\Theta$, $\hat{S}_{R_i}$ ($\hat{A}_{C_i}$) denotes the (anti-)symmetrization of the slots corresponding to the entries in the $i$-th row (column) of the tableau $\Theta$, and $\alpha_\theta$ is the normalization factor determined by the idempotency $Y_\Theta^2 = Y_\Theta$.

Practically, $\alpha_\theta$ is calculated as follows: Since $\hat{S}_{R_i}$ and $\hat{A}_{C_i}$ also have the normalization factors determined by the idempotency, $\alpha_\theta$ is given by the product of the normalization factors of all $\hat{S}_{R_i}$ and $\hat{A}_{C_i}$ over the product of the hook lengths of all the boxes of $\theta$, say $\| \theta \|$.
For instance, let $\Theta$ be the Young tableau
$Y_{{\tiny \begin{ytableau} a & b\\ c & d \end{ytableau}}}\:$.
Then, $Y_\Theta$ reads $Y_{{\tiny \begin{ytableau} a & b\\ c & d \end{ytableau}}}\:
=
\ytableausetup{boxsize=3pt}
\alpha_{{\tiny \begin{ytableau} \: & \: \\ \: & \: \end{ytableau}}}\:
\hat{A}_{ac} \hat{A}_{bd} \hat{S}_{cd} \hat{S}_{ab}$
where $\alpha_{{\tiny \begin{ytableau} \: & \: \\ \: & \: \end{ytableau}}}\:
= \ytableausetup{boxsize=6pt}(2!)^4/\| {\tiny \begin{ytableau} \: & \: \\ \: & \: \end{ytableau}} \| = 4/3$.
Especially, $Y_{{\tiny \begin{ytableau} a & b\\ c & d \end{ytableau}}}$
projects a tensor field of order $4$ onto the part of the representation
${\tiny \begin{ytableau} \: & \:\\ \: & \: \end{ytableau}}\:$,
which has the same representation as the Riemann curvature tensor.

\subsection{Prolonged equations}
In this subsection, we first provide the prolonged equations for the Killing equation without any proof. A sketch of proof is shown later in this subsection.

To provide the prolonged equations, we introduce the prolongation variables,
\begin{align}
&K^{(q)}_{b_{q}\cdots b_1a_p\cdots a_1} ~\equiv~
Y_{{\tiny \begin{ytableau} a_1 &\none[...] & \none[...] & \none[...] &a_p \\
		b_1 & \none[...] & b_q
		\end{ytableau}}}~
\nabla_{b_q \cdots b_1} K_{a_p \cdots a_1}\:,
&& (1\leq q\leq p)
\label{eq:prolong_variables}
\end{align}
where $\nabla_{ab\cdots c}\equiv \nabla_a\nabla_{b} \cdots \nabla_c$, $K_{a_p \cdots a_1}$ is a KT of order $p$.
We remark that one needs $(p+1)$ prolongation variables
to carry out the prolongation for KTs,
while that for Killing-Yano tensor fields involves
only two prolongation variables for any order
(see \cite{Semmelmann:2002,Houri:2015} or \ref{app:Yano}).
This fact complicates the prolongation for the Killing equation \eqref{eq:Killing}.

We are now ready to provide the prolonged equations. 
The prolongation for the Killing equation of order $p$ can be achieved as follows:
\begin{align}
\nabla_c K_{a_p \cdots a_1} ~=&~
Y_{{\tiny \begin{ytableau} a_1 &\none[...] &a_p
		\end{ytableau}}}~K^{(1)}_{c a_p \cdots a_1}\:,
\label{eq:prolong_p_0th}\\
\nabla_c K^{(q)}_{b_q \cdots b_1a_p\cdots a_1}
~=&~
Y_{{\tiny \begin{ytableau} a_1 &\none[...] &\none[...] &\none[...] &a_p\\
		b_1 & \none[...] & b_q
		\end{ytableau}}}~
\biggl(
\Bigl[
Y_{{\tiny \begin{ytableau} a_1 &\none[...] &\none[...] &\none[...] &a_p\\
		b_1 & \none[...] & b_q \\
		c
		\end{ytableau}}}
+Y_{{\tiny \begin{ytableau} a_1 &\none[...] &\none[...] &\none[...] &a_p & c\\
		b_1 & \none[...] & b_q
		\end{ytableau}}}
+\sum^{q}_{i=2} Y_{{\tiny \begin{ytableau} a_1
		&\none[...]&\none[...] &\none[...] &\none[...] &\none[...] &\none[...] &a_p & b_i\\
		b_1 & \none[...] & \Slash{b}_i& \none[...] & b_q & c
		\end{ytableau}}}~\Bigr]
\nabla_{c b_q \cdots b_1} K_{a_p \cdots a_1} \notag\\
&+K^{(q+1)}_{c b_q \cdots b_1 a_p\cdots a_1}\biggr)\:,
\qquad \qquad \qquad \qquad \qquad (1\leq q\leq p-1)
\label{eq:prolong_p_qth}\\
\nabla_c K^{(p)}_{b_{p} \cdots b_1a_p\cdots a_1}
~=&~
Y_{{\tiny \begin{ytableau}
		a_1 &\none[...] &a_p\\
		b_1 & \none[...] & b_p
		\end{ytableau}}}~
\Bigl[
Y_{{\tiny \begin{ytableau}
		a_1 &\none[...] &a_p\\
		b_1 & \none[...] & b_p\\
		c
		\end{ytableau}}}~
+Y_{{\tiny \begin{ytableau}
		a_1 &\none[...] &a_p & c\\
		b_1 & \none[...] & b_p
		\end{ytableau}}}~
+\sum^{p}_{i=2}
Y_{{\tiny \begin{ytableau}
		a_1 &\none[...] &\none[...] &\none[...] &\none[...] &a_p&b_i\\
		b_1 & \none[...] &\Slash{b}_i &\none[...]& b_p & c
		\end{ytableau}}}~
\Bigr]
\nabla_{c b_p \cdots b_1} K_{a_p \cdots a_1}\:,
\label{eq:prolong_p_pth}
\end{align}
where the slashed index $\Slash{b}_i$ is deleted from the Young tableau.

It is noteworthy to comment that the derivative terms look like being left on the right-hand side. However, by virtue of the properties of Young symmetrizers,
those terms can be replaced with non-derivative terms whose coefficients consist of the Riemann curvature tensor and its derivatives.
The proof is given by induction with respect to $q$ as follows:
For a fixed $q$ ($1\leq q\leq p$), the first term in the parenthesis of Eqs.\ \eqref{eq:prolong_p_qth} and \eqref{eq:prolong_p_pth} reads
\begin{align}
Y_{{\tiny \begin{ytableau} a_1 &\none[...] &\none[...] &\none[...] &a_p\\
		b_1 & \none[...] & b_q \\
		c
		\end{ytableau}}}~
\nabla_{cb_q \cdots b_1} K_{a_p \cdots a_1}
\propto&~\hat{A}_{a_q b_q}~ \cdots\hat{A}_{a_2 b_2}~
\bigl( \hat{A}_{a_1b_1 c} \nabla_{c (b_q\cdots b_1)} K_{a_p \cdots a_1} \bigr)\:.
\label{eq:derivative_proof0}
\end{align}
Performing the symmetrization over the indices $b_1,\cdots,b_q$ in the above expression, we obtain the $q!$ terms.
For each term, we then exchange $b_1$ with
the index immediately to the left repeatedly as
\begin{align*}
\nabla_{cb_q \cdots b_2 b_1} ~=~ \nabla_{cb_q \cdots b_1 b_2} + \nabla_{cb_q \cdots [b_2 b_1]}
~=~\nabla_{cb_q \cdots b_1 b_3 b_2}+\nabla_{cb_q \cdots [b_3 b_1] b_2} + \nabla_{cb_q \cdots [b_2 b_1]} ~=~ \cdots\:,
\end{align*}
until $b_1$ comes next to $c$.
After that, we act $\hat{A}_{a_1b_1 c}$ on the resulting terms
so as to replace the outer two derivatives $\nabla_{cb_1}$ with the Riemann curvature tensor,
confirming that Eq.\ \eqref{eq:derivative_proof0} can be cast in the prolongation variables of lower orders than $q$
with the coefficients of the Riemann curvature tensor and its derivatives.
Similarly,
the summands in Eqs.\ \eqref{eq:prolong_p_qth} and \eqref{eq:prolong_p_pth} can read
\begin{align*}
Y_{{\tiny \begin{ytableau} a_1
		&\none[...]&\none[...] &\none[...] &\none[...] &a_p & b_2\\
		b_1 & \none[...] & b_q & c
		\end{ytableau}}}~
\nabla_{c b_{q} \cdots b_1} K_{a_p \cdots a_1}
&~=~ Y_{{\tiny \begin{ytableau} a_1
		&\none[...]&\none[...] &\none[...] &\none[...] &a_p & b_2\\
		b_1 & \none[...] & b_q & c
		\end{ytableau}}}~
2\nabla_{c b_{q} \cdots [b_2 b_1]} K_{a_p \cdots a_1} \:,
\\
Y_{{\tiny \begin{ytableau} a_1
		&\none[...]&\none[...] &\none[...] &\none[...] &a_p & b_3\\
		b_1 & \none[...] & b_q & c
		\end{ytableau}}}~
\nabla_{c b_{q} \cdots b_1} K_{a_p \cdots a_1}
&~=~
Y_{{\tiny \begin{ytableau} a_1
		&\none[...]&\none[...] &\none[...] &\none[...] &a_p & b_3\\
		b_1 & \none[...] & b_q & c
		\end{ytableau}}}~
\bigl(2 \nabla_{c b_{q} \cdots [b_3 b_2] b_1} K_{a_p \cdots a_1}
+2 \nabla_{c b_{q} \cdots b_2 [b_3 b_1]} K_{a_p \cdots a_1}
\bigr)\:,
\end{align*}
and so on.
We deduce that all the summands can also be cast in the prolongation variables of lower orders than $q$
with the coefficients of the Riemann curvature tensor and its derivatives.
We therefore conclude that Eqs.\ \eqref{eq:prolong_p_0th}--\eqref{eq:prolong_p_pth} are
sufficient to state that the prolongation has been completed.

We show the explicit forms of the prolonged equations for $p=1$, $2$ and $3$.
The derivation of these results can be found in 
\ref{app:prolong}:
the prolonged equations for KVs are given by
	\begin{align}
	\nabla_{b} K_a ~=&~ K_{ba}^{(1)}\:,
	\label{eq:prolong_10}\\
	\nabla_{c} K_{ba}^{(1)} ~=&~
	Y_{{\tiny \begin{ytableau} a\\ b
			\end{ytableau}}}~
	Y_{{\tiny \begin{ytableau} a & c \\ b
			\end{ytableau}}}~R_{cba}\:^d K_{d} \:,
	\label{eq:prolong_11}
	\end{align}
which completely agree with Eqs.\ \eqref{eq:prolong_1pre0} and \eqref{eq:prolong_1pre1};
for order $2$, the prolonged equations are given by
	\begin{align}
	\nabla_{c} K_{ba} ~=&~
	Y_{{\tiny \begin{ytableau} a & b\end{ytableau}}}~K_{cba}^{(1)}\:,
	\label{eq:prolong_20}\\
	\nabla_{d} K_{cba}^{(1)}
	~=&~
	Y_{{\tiny \begin{ytableau} a & b \\ c\end{ytableau}}}~
	\Bigl[ K^{(2)}_{dcba}
	- \tfrac{5}{2} R_{dac}\:^mK_{mb}
	- 2 R_{dab}\:^mK_{mc}
	+\tfrac{1}{2} R_{acb}\:^mK_{md}
	\Bigr] \:,
	\label{eq:prolong_21}\\
	\nabla_{e} K_{dcba}^{(2)}
	~=&~
	Y_{{\tiny \begin{ytableau} a & b \\ c & d\end{ytableau}}}~
	\Bigl[ -\tfrac{4}{3} (\nabla_a R_{bcd}\:^m) K_{me}
	-\tfrac{2}{3} (\nabla_e R_{cab}\:^m) K_{md}
	-\tfrac{8}{3} (\nabla_a R_{bde}\:^m) K_{mc}
	-12 R_{eac}\:^m K_{mdb}^{(1)} \notag\\
	&-4 R_{eab}\:^m K_{mcd}^{(1)}
	-\tfrac{2}{3} R_{cab}\:^m K_{mde}^{(1)}
	+\tfrac{7}{3} R_{cab}\:^m K_{med}^{(1)}
	\Bigr]\:.
	\label{eq:prolong_22}
	\end{align}
Compared with the results of \cite{Veblen:1923,Hauser:1974,Hauser:1975i,Hauser:1975ii}, our results have simpler forms;
taking one more step, we can write out the prolonged equations
for order $3$ explicitly.
	\begin{align}
	\nabla_d K_{cba} ~=&~
	Y_{{\tiny \begin{ytableau} a & b & c\end{ytableau}}}~
	K^{(1)}_{dcba}\:,
	\label{eq:prolong_30}\\
	\nabla_e K^{(1)}_{dcba} ~=&~
	Y_{{\tiny \begin{ytableau} a & b & c \\ d\end{ytableau}}}~
	\Bigl[
	K^{(2)}_{edcba}
	-3R_{ead}\:^m K_{mbc}
	-5R_{eab}\:^m K_{mdc}
	-R_{dab}\:^m K_{mce}
	\Bigr] \:,
	\label{eq:prolong_31}\\
	\nabla_{f} K^{(2)}_{edcba} ~=&~
	Y_{{\tiny \begin{ytableau} a & b & c \\ d & e\end{ytableau}}}~
	\Bigl[
	K^{(3)}_{fedcba}
	+2(\nabla_b R_{ead}\:^m) K_{mcf}
	+2(\nabla_b R_{eac}\:^m) K_{mdf}
	+2(\nabla_b R_{fac}\:^m) K_{mde} \notag \\
	&
	+6(\nabla_b R_{fad}\:^m) K_{mce}
	-2(\nabla_b R_{fda}\:^m) K_{mce}
	-\tfrac{32}{3} R_{fbc}\:^m K^{(1)}_{meda}
	-\tfrac{16}{3} R_{fba}\:^m K^{(1)}_{mced} 	\notag\\
	&
	-20 R_{fbe}\:^m K^{(1)}_{mcda}
	+\tfrac{2}{3}R_{cea}\:^m K^{(1)}_{mbdf}
	-\tfrac{10}{3}R_{bea}\:^m K^{(1)}_{mfcd}
	+\tfrac{2}{3}R_{bea}\:^m K^{(1)}_{mfdc}
	\Bigr] \:,
	\label{eq:prolong_32}\\
	\nabla_{g} K^{(3)}_{fedcba} ~=&~
	Y_{{\tiny \begin{ytableau} a & b & c \\ d & e & f\end{ytableau}}}~
	\Bigl[
	-24 R_{gfc}\:^m K^{(2)}_{mdbea}
	-6 R_{gfd}\:^m K^{(2)}_{mecba}
	+4 R_{fcd}\:^m K^{(2)}_{mgbae}
	-12 R_{fcd}\:^m K^{(2)}_{mgeba} \notag \\
	&
	-20 (\nabla_d R_{aeg}\:^m) K^{(1)}_{mbcf}
	+12 (\nabla_d R_{age}\:^m) K^{(1)}_{mbcf}
	-2 (\nabla_f R_{gde}\:^m) K^{(1)}_{mcba}\notag \\
	&
	-\tfrac{3}{2} (\nabla_d R_{aef}\:^m) K^{(1)}_{mbcg}
	-16 (\nabla_d R_{aeb}\:^m) K^{(1)}_{mgcf}
	-3 (\nabla_d R_{aef}\:^m) K^{(1)}_{mgbc} \notag \\
	&
	+\tfrac{9}{2} (\nabla_{fe} R_{bda}\:^m) K_{mgc}
	-\tfrac{9}{2} (\nabla_{ed} R_{fcg}\:^m) K_{mba}
	+3 (\nabla_{ge} R_{afc}\:^m) K_{mdb}\notag \\
	&
	+6 R_{gfc}\:^m R_{ebd}\:^n K_{mn a}
	+5 R_{gfe}\:^m R_{dac}\:^n K_{mn b}
	+6 R_{gfe}\:^m R_{mac}\:^n K_{ndb} \notag \\
	&
	-4 R_{fbe}\:^m R_{mgd}\:^n K_{nac}
	+ \tfrac{1}{2} R_{fbe}\:^m R_{mdg}\:^n K_{nac}
	-9 R_{dbf}\:^m R_{mcg}\:^n K_{nea} \notag \\
	&
	-2 R_{fce}\:^m R_{mbd}\:^n K_{n ag}
	+\tfrac{11}{2} R_{fce}\:^m R_{mdb}\:^n K_{n ag}
	\Bigr]\:.
	\label{eq:prolong_33}
	\end{align}

Let us provide a sketch of the proof for the results \eqref{eq:prolong_p_0th}--\eqref{eq:prolong_p_pth}.
Since $K_{a_p\cdots a_1}$ is totally symmetric, we have
\begin{align*}
\nabla_c K_{a_p\cdots a_1}
~=~
	Y_{{\tiny \begin{ytableau} a_1 & \none[...] & a_p \end{ytableau}}}~
	\mathrm{id}_{p+1}\:
	\nabla_c K_{a_p\cdots a_1} \:,
\end{align*}
where $\mathrm{id}_{p+1}$ is the identity operator.
Using the completeness of the Young symmetrizers with Littlewood's correction \eqref{def:littlewood} yields
\begin{align*}
&Y_{{\tiny \begin{ytableau} a_1 & \none[...] & a_p \end{ytableau}}}~
\mathrm{id}_{p+1}\:
\nabla_c K_{a_p\cdots a_1}
=~ Y_{{\tiny \begin{ytableau} a_1 & \none[...] & a_p \end{ytableau}}}~
\Big(
L_{{\tiny \begin{ytableau} a_1 & \none[...] & a_p \\ c\end{ytableau}}}~+\cdots
\Big)
\nabla_c K_{a_p\cdots a_1} \:.
\end{align*}
The round brackets contain a lot of the Young symmetrizers.
However, most of these symmetrizers vanish 
due to Pieri's formula \eqref{eq:pieri} and the Killing equation \eqref{eq:Killing},
leaving only $L_{{\tiny \begin{ytableau} a_1 & \none[...] & a_p \\ c\end{ytableau}}}\:$.
Thus we obtain
\begin{align}
\nabla_c K_{a_p\cdots a_1}
~=~ Y_{{\tiny \begin{ytableau} a_1 & \none[...] & a_p \end{ytableau}}}~
L_{{\tiny \begin{ytableau} a_1 & \none[...] & a_p \\ c\end{ytableau}}}~
\nabla_cK_{a_p\cdots a_1} \:.
\label{eq:formula_0}
\end{align}
The tableau ${\tiny \begin{ytableau} a_1 & \none[...] & a_p \\ c\end{ytableau}}\:$ is row-ordered
and then $L_{{\tiny \begin{ytableau} a_1 & \none[...] & a_p \\ c\end{ytableau}}}\:$ is equal to $Y_{{\tiny \begin{ytableau} a_1 & \none[...] & a_p \\ c\end{ytableau}}}\:$,
confirming Eq.\ \eqref{eq:prolong_p_0th}.
Similarly, differentiating the $q$th prolongation variable \eqref{eq:prolong_variables} for $1\leq q\leq p$ gives
\begin{align}
&\nabla_c K^{(q)}_{b_q\cdots b_1a_p \cdots a_1}
~=~
Y_{{\tiny \begin{ytableau} a_1 & \none[...] &\none[...] & \none[...]& a_p \\
b_1 & \none[...] &b_q \end{ytableau}}}~
\nabla_{c b_q \cdots b_1} K_{a_p \cdots a_1} \notag\\
=&~
Y_{{\tiny \begin{ytableau} a_1 & \none[...] &\none[...] & \none[...]& a_p \\
		b_1 & \none[...] &b_q \end{ytableau}}}~
\Bigl[
L_{{\tiny \begin{ytableau} a_1 & \none[...] &\none[...] &\none[...]&\none[...]& a_p \\
		b_1 & \none[...] &b_q & c\end{ytableau}}}~
+L_{{\tiny \begin{ytableau} a_1 & \none[...] &\none[...] & \none[...]& a_p \\
		b_1 & \none[...] &b_q \\ c\end{ytableau}}}~
+L_{{\tiny \begin{ytableau} a_1 & \none[...] &\none[...] & \none[...]& a_p& c \\
		b_1 & \none[...] &b_q \end{ytableau}}} \label{eq:proof}
+\sum^{q}_{i=2} L_{{\tiny \begin{ytableau} a_1
		&\none[...]&\none[...] &\none[...] &\none[...] &\none[...] &\none[...] &a_p & b_i\\
		b_1 & \none[...] & \Slash{b}_i& \none[...] & b_q & c
		\end{ytableau}}}~
\Bigr]
\nabla_{cb_q \cdots b_1} K_{a_p \cdots a_1}\:.
\end{align}
As we see from Eq.\ \eqref{eq:formula_102}--\eqref{eq:formula_103},
all Littlewood's corrections in the above expression vanish
and thus $L_\Theta$ equals $Y_\Theta$.
We therefore obtain Eq.\ \eqref{eq:prolong_p_qth},
concluding the proof.
Note that the expression \eqref{eq:proof} is also valid for $q=p$
if $L_{{\tiny \begin{ytableau} a_1 & \none[...] &\none[...]&\none[...]&\none[...]& a_p \\
b_1 & \none[...] &b_q & c\end{ytableau}}}\:$
is omitted.

\subsection{Geometric interpretation}

Once the prolonged equations \eqref{eq:prolong_p_0th}--\eqref{eq:prolong_p_pth} have been formulated,
one may forget the definitions of the prolongation variables \eqref{eq:prolong_variables} because one can reconstruct Eqs.\ \eqref{eq:Killing} and \eqref{eq:prolong_variables} from the prolonged equations
\eqref{eq:prolong_p_0th}--\eqref{eq:prolong_p_pth} under the assumption
\begin{align}
K^{(q)}_{b_q \cdots b_1a_p\cdots a_1} ~=~
Y_{{\tiny \begin{ytableau} a_1 &\none[...] & \none[...] & \none[...] &a_p \\
		b_1 & \none[...] & b_q
		\end{ytableau}}}~
K^{(q)}_{b_q \cdots b_1a_p\cdots a_1} \:,
\label{eq:assumption}
\end{align}
which means that the prolonged equations \eqref{eq:prolong_p_0th}--\eqref{eq:prolong_p_pth}
with the assumption \eqref{eq:assumption} are equivalent to the Killing equation \eqref{eq:Killing}.
A proof of this assertion is given as follows:
Suppose the prolonged equations \eqref{eq:prolong_p_0th}--\eqref{eq:prolong_p_pth}
with the assumption \eqref{eq:assumption} hold.
First, multiplying both sides of Eq.\ \eqref{eq:prolong_p_0th} by
$Y_{{\tiny \begin{ytableau} a_1 & \none[...] & a_p & c\end{ytableau}}}\:$
from the left gives
\begin{align}
\nabla_{(c} K_{a_p \cdots a_1)}
~=~
Y_{{\tiny \begin{ytableau} a_1 & \none[...] & a_p & c\end{ytableau}}}~
Y_{{\tiny \begin{ytableau} a_1 & \none[...] & a_p \\ c\end{ytableau}}}~
K^{(1)}_{c a_p\cdots a_1} ~=~
0\:,
\end{align}
confirming the Killing equation \eqref{eq:Killing}.
We have used the orthogonality of Young symmetrizers \eqref{eq:orthogonality} here.
Next, multiplying both sides of Eq.\ \eqref{eq:prolong_p_qth} by
$Y_{{\tiny \begin{ytableau} a_1 & \none[...] & \none[...] & \none[...]& \none[...] & a_p \\
		b_1 & \none[...] & b_q & c \end{ytableau}}}\:$
from the left yields
\begin{align}
Y_{{\tiny \begin{ytableau} a_1 & \none[...] & \none[...] & \none[...]& \none[...] & a_p \\
		b_1 & \none[...] & b_q & c \end{ytableau}}}~
\nabla_c K^{(q)}_{b_q \cdots b_1 a_p\cdots a_1}
&=~
Y_{{\tiny \begin{ytableau} a_1 & \none[...] & \none[...] & \none[...]& \none[...] & a_p \\
		b_1 & \none[...] & b_q & c \end{ytableau}}}~
Y_{{\tiny \begin{ytableau} a_1 & \none[...] & \none[...] & \none[...]& \none[...]& a_p \\
		b_1 & \none[...] & b_q \end{ytableau}}}~
K^{(q+1)}_{c b_q \cdots b_1a_p\cdots a_1}
~=~
K^{(q+1)}_{c b_q \cdots b_1a_p\cdots a_1}\:,
\end{align}
which leads to
\begin{align}
K^{(q+1)}_{c b_q \cdots b_1a_p\cdots a_1}
&=~ Y_{{\tiny \begin{ytableau} a_1 & \none[...] & \none[...] & \none[...]& \none[...] & a_p \\
		b_1 & \none[...] & b_q & c \end{ytableau}}}~
\cdots Y_{{\tiny \begin{ytableau} a_1 & \none[...] & a_p \\
		b_1 \end{ytableau}}}~
\nabla_{c b_q \cdots b_1} K_{a_p \cdots a_1}
~=~ Y_{{\tiny \begin{ytableau} a_1 & \none[...] & \none[...] & \none[...]& \none[...] & a_p \\
		b_1 & \none[...] & b_q & c \end{ytableau}}}~
\nabla_{c b_q \cdots b_1} K_{a_p \cdots a_1}\:,
\end{align}
where we have used the identity
\begin{align}
Y_{{\tiny \begin{ytableau} a_1 & \none[...] & \none[...] & \none[...]& \none[...] & a_p \\
		b_1 & \none[...] & b_i & c\end{ytableau}}}~
Y_{{\tiny \begin{ytableau} a_1 & \none[...] & \none[...] & \none[...]& \none[...] & a_p \\
		b_1 & \none[...] & b_i \end{ytableau}}}~
~=~Y_{{\tiny \begin{ytableau} a_1 & \none[...] & \none[...] & \none[...]& \none[...] & a_p \\
		b_1 & \none[...] & b_i & c\end{ytableau}}}\:,
\end{align}
which follows from Schur's lemma \eqref{eq:schur} and
Raicu's formula \eqref{eq:raicu}.

\ytableausetup{boxsize=6.5pt}
Geometrically, the set of the variables \eqref{eq:assumption} can be viewed as a section of the vector bundle $E^{(p)}$ over $M$
\begin{align}
E^{(p)} ~=~
\underbrace{{\tiny
		\begin{ytableau} \quad &\quad &\quad &\none[...] &\quad
		\end{ytableau}}}_{p\:\:\text{boxes}}
~\oplus
\underbrace{{\tiny
		\begin{ytableau} \quad &\quad &\quad &\none[...] &\quad \\
								\quad 
		\end{ytableau}}}_{(p+1)\:\:\text{boxes}}
\oplus~ \cdots
~\oplus~
\underbrace{{\tiny
		\begin{ytableau} \quad &\quad &\quad &\none[...] &\quad \\
								\quad &\quad &\quad &\none[...] &\quad
		\end{ytableau}}}_{2p\:\:\text{boxes}}~,
\end{align}
where the fibers are irreducible representations of GL$(n)$
corresponding to the Young diagrams.
Moreover, the prolonged equations
\eqref{eq:prolong_p_0th}--\eqref{eq:prolong_p_pth} can be viewed as the parallel equation
for a section of $E^{(p)}$,
\begin{align}
D_a {\boldsymbol K} ~=~ 0\:,
\label{eq:parallel_eq}
\end{align}
where $D_a \equiv \nabla_a - \Omega_a$
is the connection on $E^{(p)}$ and
${\boldsymbol K}$ is a section of $E^{(p)}$.
$\Omega_a \in \textrm{End}(E^{(p)})$ depends on the Riemann curvature tensor and its derivatives up to
$(p-1)$th order which can be read off from
the right-hand side of the prolonged equations \eqref{eq:prolong_p_0th}--\eqref{eq:prolong_p_pth}.
Hence it turns out that there is a one-to-one correspondence
between KTs of order $p$ and the parallel sections.

\section{Integrability conditions}\label{sec:int_cond}

This section is devoted to investigating
the integrability condition of 
the $q$th prolongation variable of a KT of order $p$,
which arises as a consistency condition:
\begin{align}
0 ~=~
2\:I_{a_1 \cdots a_p b_1 \cdots b_q cd}^{(p,q)} ~\equiv~
\nabla_{dc}K^{(q)}_{b_q \cdots b_1a_p\cdots a_1}
-\nabla_{cd}K^{(q)}_{b_q \cdots b_1a_p\cdots a_1}
-2\nabla_{[dc]}K^{(q)}_{b_q \cdots b_1a_p\cdots a_1}\:,
\label{eq:tensor_i}
\end{align}
where the first two terms in Eq.\ \eqref{eq:tensor_i}
are evaluated by the $q$th prolonged equation \eqref{eq:prolong_p_qth};
on the one hand, the last term in Eq.\ \eqref{eq:tensor_i}
is described by the defining equation of the Riemann curvature tensor.

\subsection{Main results}
Calculating the integrability condition \eqref{eq:tensor_i} up to $p=3$, we obtain
the following results:

\ytableausetup{boxsize=6pt}
\underline{$p=1$}
	\begin{align}
	I_{abc}^{(1,0)} ~=&~ 0 \:, \label{eq:int_10}\\
	I_{abcd}^{(1,1)} ~=&~
	Y_{{\tiny \begin{ytableau} a & c\\
	b & d \end{ytableau}}}~
	\left[ (\nabla_d R_{cba}\:^m) K_m - 2 R_{cba}\:^m K^{(1)}_{m d}\right]\:,
	\label{eq:int_11}
	\end{align}

\underline{$p=2$}
	\begin{align}
	I_{abcd}^{(2,0)} ~=&~ I_{abcde}^{(2,1)} = 0\:, \label{eq:int_20}\\
	I_{abcdef}^{(2,2)} ~=&~
	Y_{{\tiny \begin{ytableau} a & b & e\\
			c & d & f\end{ytableau}}}~
	\Bigl[
	3 (\nabla_{fd} R_{ecb}\:^m) K_{ma}
	+2 R_{edc}\:^{m} R_{m bf}\:^{n} K_{na}
	-5 R_{edc}\:^{m} R_{m fb}\:^{n} K_{na}\notag\\
	&+3 (\nabla_f R_{eda}\:^m) K^{(1)}_{m bc}
	-9 (\nabla_f R_{eda}\:^m) K^{(1)}_{m cb}
	- 8 R_{fed}\:^m K^{(2)}_{m cba}
	\Bigr]\:,
	\label{eq:int_22}
	\end{align}

\underline{$p=3$}
	\begin{align}
	I_{abcde}^{(3,0)} ~=&~ I_{abcdef}^{(3,1)} = I_{abcdefg}^{(3,2)}= 0\:, \label{eq:int_30}\\
	I_{abcdefgh}^{(3,3)} ~=&~
	Y_{{\tiny \begin{ytableau} a & b & c & g\\
			d & e & f & h\end{ytableau}}}~
	\Bigl[
	6(\nabla_{hfe} R_{gdc}\:^m) K_{mba}
	-27 (\nabla_{h} R_{gfe}\:^m) R_{mdc}\:^n K_{n ba}
	-34 R_{ghf}\:^m (\nabla_e R_{mdc}\:^n) K_{nba} \notag\\
	&-15 (\nabla_h R_{gfe}\:^m) R_{mc b}\:^n K_{nda}
	+15 (\nabla_h R_{gfe}\:^m) R_{dcb}\:^n K_{mn a}
	-20 R_{ghf}\:^m (\nabla_{e} R_{mcb}\:^n) K_{nda} \notag \\
	&+20 R_{ghf}\:^m (\nabla_{e} R_{dcb}\:^n) K_{mn a}
	-24 (\nabla_{hf} R_{ceg}\:^m) K^{(1)}_{mbad}
	+12 (\nabla_{hf} R_{ceg}\:^m) K^{(1)}_{mdba} \notag\\
	&+50 R_{ghf}\:^m R_{mec}\:^n K^{(1)}_{n bad}
	+40 R_{ghf}\:^m R_{mce}\:^n K^{(1)}_{n dba}
	-\tfrac{74}{3} R_{ghf}\:^mR_{med}\:^n K^{(1)}_{n cba} \notag \\
	&-\tfrac{40}{3} R_{ghf}\:^m R_{ced}\:^n K^{(1)}_{mn ba} 
	-35 (\nabla_h R_{fgc}\:^m) K^{(2)}_{mbaed}
	- 5 (\nabla_{h} R_{fgc}\:^m) K^{(2)}_{medba}\notag\\
	&-20 R_{hgf}\:^m K^{(3)}_{medcba}
	\Bigr]\:.
	\label{eq:int_33}
	\end{align}

From Eqs.\ \eqref{eq:int_10}--\eqref{eq:int_33}, it is observed that
in the cases $q<p$ the integrability condition of $q$th prolongation variable
is automatically satisfied.
More precisely, we can confirm that all of these conditions vanish identically,
up to the first and second Bianchi identities, $R_{[abc]}\:^d = \nabla_{[a} R_{bc]de} = 0$.
In contrast, the integrability condition at $q=p$ provides
nontrivial relations among all the prolongation variables.
This is consistent with the result in \cite{Kruglikov:2016}.

It is intriguing to note that the integrability condition of the $p$th prolongation variable belongs to
the Young diagram of shape $(p+1, p+1)$.
If we act the curvature operator on the $p$th prolongation variable,
we obtain a $(2p+2)$th order tensor
belonging to the representation $(1,1) \otimes (p,p)$.
It can be decomposed into the irreducible representations $(p,p,1,1)$, $(p+1,p,1)$ and $(p+1,p+1)$
that are respectively described by the Young diagrams
\begin{align*}
&{\tiny \begin{ytableau} \quad & \none[...] & \quad\\
		\quad & \none[...] & \quad\\
		\quad\\
		\quad
		\end{ytableau}}~,
&&{\tiny \begin{ytableau} \quad & \none[...] & \quad & \quad\\
	\quad & \none[...] & \quad\\
	\quad
	\end{ytableau}}~,
&&{\tiny \begin{ytableau} \quad & \none[...] & \quad & \quad\\
		\quad & \none[...] & \quad & \quad
		\end{ytableau}}~.
\end{align*}
However, we observe from Eqs.\ \eqref{eq:int_11}, \eqref{eq:int_22} and \eqref{eq:int_33}
that the integrability condition of the $p$th prolongation variable
makes non-trivial contribution only for the representation $(p+1,p+1)$.
For instance the integrability condition $I_{abcd}^{(1,1)}$
could have the representations
\begin{align*}
&{\tiny \begin{ytableau} \quad\\ \quad\\ \quad \\ \quad\end{ytableau}}~,
&&{\tiny \begin{ytableau} \quad & \quad\\ \quad \\ \quad\end{ytableau}}~,
&&{\tiny \begin{ytableau} \quad & \quad\\ \quad & \quad\end{ytableau}}~.
\end{align*}
However, the result \eqref{eq:int_11} claims that the first two representations do not appear for some reason.
Thus we are led to make the following conjecture:

\quad\\
\noindent
{\bf Conjecture.}
{\it The integrability condition of the $p$th prolongation variable belongs to the representation described by the rectanguler Young diagram $(p+1, p+1)$.
}\\

In general, it is not easy to write out the integrability condition of the $p$th prolongation variable.
This difficulty becomes more prominent as the order of KTs increases.
However, if this conjecture holds true for $p \geq 4$,
then there is no need to calculate the terms that belong to the representations
${\tiny \begin{ytableau} \quad & \none[...] & \quad\\
	\quad & \none[...] & \quad\\
	\quad\\
	\quad
	\end{ytableau}}\:$ and ${\tiny \begin{ytableau} \quad & \none[...] & \quad & \quad\\
	\quad & \none[...] & \quad\\
	\quad
	\end{ytableau}}\:$,
thereby allowing us to obtain the formula
\begin{align}
	&I_{a_1 \cdots a_p b_1 \cdots b_p cd}^{(p,p)} ~=~
	Y_{{\tiny \begin{ytableau} a_1 & \none[...] & a_p & c\\
			b_1 & \none[...] & b_p & d\end{ytableau}}}~\Bigl[
	2 \nabla_{d [c b_p] \cdots b_1} K_{a_p \cdots a_1}
	+3 \nabla_{d b_p [c b_{p-1}] \cdots b_1} K_{a_p \cdots a_1} \notag \\
	&+4 \nabla_{d b_p b_{p-1} [c b_{p-2}] \cdots b_1} K_{a_p \cdots a_1}
	\cdots +(p+1) \nabla_{d b_p b_{p-1} b_{p-2} \cdots [c b_1]} K_{a_p \cdots a_1}
	- \nabla_{[dc]} K^{(p)}_{b_p \cdots b_1 a_p\cdots a_1}\Bigr] \notag\\
	&+ (\text{the terms that belong to the representations}
	~{\tiny \begin{ytableau} \quad & \none[...] & \quad\\
		\quad & \none[...] & \quad\\
		\quad\\
		\quad
		\end{ytableau}}~ \text{and}
	~{\tiny \begin{ytableau} \quad & \none[...] & \quad & \quad\\
		\quad & \none[...] & \quad\\
		\quad
		\end{ytableau}}~
	) \:. \label{eq:int_pp}
\end{align}
We have confirmed that for the cases $p \leq 3$,
the last term exactly vanish up to the first and second Bianchi identities, $R_{[abc]}\:^d = \nabla_{[a} R_{bc]de} = 0$.
The proof of the formula \eqref{eq:int_pp} is given by \ref{app:d_int}.

As is the case with the prolonged equations \eqref{eq:prolong_p_0th}--\eqref{eq:prolong_p_pth},
there are still a lot of derivative terms left in the right-hand side of Eq.\ \eqref{eq:int_pp}.
Then again, we can rewrite all these terms in Eq.\ \eqref{eq:int_pp}
to non-derivative terms by using the prolonged equations.
For $p \geq 4$, this is a challenging and daunting task
which is beyond our scope here
and will be considered in the future.

\subsection{Application}
As an application of the integrability conditions, 
we show a method for computing
the number of linearly independent solutions
to the Killing equation.

Let us recall the parallel equation \eqref{eq:parallel_eq}.
We introduce the curvature of the connection $D_a$ as
$R_{ab}^D {\boldsymbol K} \equiv [D_a, D_b] {\boldsymbol K}$.
We call this the {\it Killing curvature}.
All the integrability conditions of a KT of order $p$ can be collectively expressed as
\begin{align}
R_{ab}^D {\boldsymbol K}~=~ 0\:.
\label{eq:int_cond_re}
\end{align}
By repeatedly differentiating the condition \eqref{eq:int_cond_re},
we obtain the set of linear algebraic equations
\begin{align}
&R_{ab}^D {\boldsymbol K}~=~ 0\:,
&&(D_a R_{bc}^D) {\boldsymbol K} ~=~ 0\:,
&&(D_a D_b R_{cd}^D) {\boldsymbol K} ~=~ 0\:,
&&\cdots
\end{align}
After working out $r$ differentiations,
we are led to the system
\begin{align}
{\boldsymbol R}^D_r {\boldsymbol K} ~=~ 0\:,
\label{eq:whole_int_cond}
\end{align}
where the coefficient matrix ${\boldsymbol R}^D_r$
depends on the Killing curvature and its derivatives.
For example,
\begin{align}
&{\boldsymbol R}^D_0 =
\begin{pmatrix}
R_{ab}^D
\end{pmatrix}\:,
&&{\boldsymbol R}^D_1 =
\begin{pmatrix}
R_{ab}^D \\
D_a R_{bc}^D
\end{pmatrix}\:,
&&{\boldsymbol R}^D_2 =
\begin{pmatrix}
R_{ab}^D \\
D_a R_{bc}^D \\
D_a D_b R_{cd}^D
\end{pmatrix}\:.
\end{align}

It is known that by applying the Frobenius theorem to the condition \eqref{eq:whole_int_cond},
the following theorem holds true (see, e.g. Ref.\ \cite{Bryant:2008}).

\quad\\
\noindent
{\bf Theorem.~}
{\it If we find the smallest natural number $r_0$ such that
\begin{align}
\mathrm{rank}\:{\boldsymbol R}^D_{r_0}
~=~ \mathrm{rank}\:{\boldsymbol R}^D_{r_0+1}\:,
\label{eq:r_0}
\end{align}
then it follows that 
$\mathrm{rank}\:{\boldsymbol R}^D_{r_0}
=\mathrm{rank}\:{\boldsymbol R}^D_{r_0+r}$
for any natural number $r$ and
the dimension of the space of the KT reads
\begin{align}
\dim K^p
~=~ \mathrm{rank}\:E^{(p)}
-\mathrm{rank}\:{\boldsymbol R}^D_{r_0}
\:,
\label{eq:theorem_det_num}
\end{align}
where $\mathrm{rank}\:E^{(p)}$
is given by the BDTT formula \eqref{eq:dim_of_KS}.
}

The condition \eqref{eq:r_0} means that
the components of the $(r_0+1)$th order derivatives of the Killing curvature,
$D_{a_1}\cdots D_{a_{r_0+1}} R^D_{bc}$,
can be expressed as the linear combinations of the components of the lower order derivatives than $r_0+1$.
Hence, the components of the one-higher order derivatives,
$D_{a_1}\cdots D_{a_{r_0+2}} R^D_{bc}$, can also be expressed as the linear combinations of the lower order derivatives than $r_0+1$.
By induction, we can conclude that the theorem holds true.

It should be remarked that
computing the rank of the matrix ${\boldsymbol R}^D_r$
boils down to solve a system of the linear algebraic equations
\begin{align}
&I_{a_1 \cdots a_p b_1 \cdots b_p cd}^{(p,p)}~=~0\:,
&& \cdots\:,
&&\left.\nabla_{e_1 \cdots e_r} I_{a_1 \cdots a_p b_1 \cdots b_p cd}^{(p,p)}
\right|_{D{\boldsymbol K}=0} ~=~0\:,
\end{align}
where $I_{a_1 \cdots a_p b_1 \cdots b_p cd}^{(p,p)}$ is 
the integrability condition of the $p$th prolongation variable of a KT of order $p$
defined by Eq.\ \eqref{eq:tensor_i}.
If $r_0$ exists, Eq.\ \eqref{eq:theorem_det_num} allows us to have the value of $\dim K^p$.
Otherwise differentiating the integrability condition \eqref{eq:int_cond_re}
reveals a large number of additional conditions.
We can stop the differentiation and conclude that no KT of order $p$ exists
if $\mathrm{rank}\:{\boldsymbol R}^D_{r_0}$ is equal to $\mathrm{rank}\:E^{(p)}$.
Based on this fact, we can determine the dimension of the space of KTs.

To demonstrate the efficacy of our method,
let us take the Kerr metric in Boyer-Lindquist coordinates:
\begin{align}
ds^2 ~=~&-\left(1-\frac{2Mr}{\Sigma}\right) dt^2
-\frac{4aMr\sin ^2\theta}{\Sigma} dt d\phi
+\frac{\Sigma}{\Delta} \:dr^2
+\Sigma \:d\theta^2 \notag\\
&+\left(r^2+a^2+\frac{2a^2Mr \sin ^2\theta}{\Sigma}\right)
\sin ^2\theta d\phi^2\:,
\label{eq:Kerr}
\end{align}
with
\begin{align}
&\Sigma ~=~ r^2+a^2 \cos ^2\theta\:,
&&\Delta ~=~ r^2 -2Mr+a^2\:,
\end{align}
and determine the number of the solutions to the Killing equation up to $p=2$.
As a higher order KT includes reducible ones, e.g.
$\xi_{(a} \zeta_{b)}$ is a trivial KT
if $\xi^a$ and $\zeta^a$ are KVs,
at first we must solve the integrability condition for $p=1$.

\begin{table}[t]
	\begin{center}
		\begin{tabular}{c|c|c}
			\hline \hline
			4-dimensional metrics / order of KTs & 1 & 2 \\
			\hline \hline
			Maximally symmetric & 10 & 50 \\
			Schwarzschild & 4 & 11 \\
			Kerr & 2 & 5 \\
			Reissner-Nordstrom & 4 & 11 \\
			\hline
		\end{tabular}
	\end{center}
	\caption{
		The number of the first and second order KTs in several regular black hole metrics.
	}
	\label{tab:number_of_KTs}
\end{table}

For $p=1$ case, a section of the bundle $E^{(1)}$
can be written as
\begin{align}
&{\boldsymbol K} ~=~
\begin{pmatrix}
K_a\\
K^{(1)}_{ba}
\end{pmatrix}\:,
&&\text{with}
&&K^{(1)}_{ba} \in {\tiny \begin{ytableau} a\\
	b\end{ytableau}}\:.
\end{align}
By solving the linear systems ${\boldsymbol R}^D_1 \:{\boldsymbol K} =0$
and ${\boldsymbol R}^D_2 \:{\boldsymbol K} = 0$, that is
\begin{align}
&I_{abcd}^{(1,1)} ~=~ 0\:,
&&\left.\nabla_e I_{abcd}^{(1,1)}\right|_{D{\boldsymbol K}=0} ~=~ 0\:,
\end{align}
and
\begin{align}
&I_{abcd}^{(1,1)} ~=~ 0\:,
&&\left.\nabla_e I_{abcd}^{(1,1)}\right|_{D{\boldsymbol K}=0} ~=~ 0\:,
&&\left.\nabla_{ef} I_{abcd}^{(1,1)}\right|_{D{\boldsymbol K}=0} ~=~ 0\:,
\label{eq:Kerr_ex:adjoin}
\end{align}
we find that $\mathrm{rank}\:{\boldsymbol R}^D_{1}
=\mathrm{rank}\:{\boldsymbol R}^D_{2} = 8$.
In other words, the adjoined equation in Eq.\ \eqref{eq:Kerr_ex:adjoin},
$\nabla_{ef} I_{abcd}^{(1,1)} =0$, does not change the rank.
Since the maximal number of the KVs is 
$\mathrm{rank} \:E^{(1)} = 10$,
we can conclude that
$\dim K^1=2$.
This is consistent with our knowledge:
the two vector fields
$\xi^a = (\partial_t)^a$
and $\zeta^a = (\partial_\phi)^a$
are the only KVs in the Kerr metric \eqref{eq:Kerr}.
Similarly, a section of the bundle $E^{(2)}$ is given by
\begin{align}
&{\boldsymbol K} ~=~
\begin{pmatrix}
K_{ba}\\
K^{(1)}_{cba}\\
K^{(2)}_{dcba}
\end{pmatrix}\:,
&&\text{with}
&&K^{(1)}_{cba} \in {\tiny \begin{ytableau} a & b\\
	c\end{ytableau}}\:,
&&K^{(2)}_{dcba} \in {\tiny \begin{ytableau} a & b\\
	c & d\end{ytableau}}\:.
\end{align}
After solving the linear systems
${\boldsymbol R}^D_1 \:{\boldsymbol K} =0$
and ${\boldsymbol R}^D_2 \:{\boldsymbol K} = 0$,
we find that $\mathrm{rank}\:{\boldsymbol R}^D_{1}
=\mathrm{rank}\:{\boldsymbol R}^D_{2} = 45$.
It also follows from Eq.\ \eqref{eq:dim_of_KS} that $\mathrm{rank} \:E^{(2)} = 50$.
This amounts to $\dim K^2= 5$.
We know that four of them
\begin{align}
&g_{ab}\:,
&&\xi_{(a}\xi_{b)}\:,
&&\xi_{(a}\zeta_{b)}\:,
&&\zeta_{(a}\zeta_{b)}\:,
\end{align}
are reducible KTs while the only one
\begin{align}
K_{ab} ~=~&
\frac{a^2}{\Sigma}\Bigl[\Delta\:\cos^2 \theta+r^2 \sin^2\theta\Bigr](dt)^2_{ab}
-\frac{a^2\:\Sigma\:\cos^2\theta}{\Delta} (dr)^2_{ab}
+r^2\:\Sigma\:(d\theta)^2_{ab} \notag\\
&
+\frac{\sin^2\theta}{\Sigma}
\Bigl[r^2(a^2+r^2)^2 +
a^4\:\Delta\:\cos^2\theta \sin^2\theta\Bigr]\:(d\phi)^2_{ab} \notag \\
&-\frac{2a \sin^2\theta}{\Sigma}\Bigl[
r^2(a^2+r^2)+a^2\Delta\:\cos^2\theta
\Bigr]\:(dt)_{(a}(d\phi)_{b)}
\:.
\end{align}

Using the same method,
we investigate the $1$st and $2$nd order KTs in several regular black hole metrics,
as shown in Table \ref{tab:number_of_KTs}.
It would be of great interest to make a systematic investigation of
higher-order KTs in various spacetimes.
We leave it as a future work.

\section{Summary and discussion}\label{sec:discussion}
In this paper we have formulated the prolonged equations and the integrability conditions of the Killing equation \eqref{eq:Killing} in a manner that uses Young symmetrizers.
In particular, we have provided the explicit forms of the prolonged equations for a general order
and the integrability conditions for $p=1$, $2$ and $3$.
We have also made a conjecture on the integrability conditions for $p\geq 4$
and shown their application for computing the dimension of the space of KTs.

While integrability conditions are simply the consequence of
the requirement that mixed partial derivatives must commute,
the explicit forms of them have brought us essential insights into general relativity, such as the Gauss–-Codazzi equations in the Hamiltonian formulation of general relativity and the Raychaudhuri equations in the derivation of the singularity theorems.
Similarly, the integrability conditions of the Killing equation
for $p=1$ lead to an immediate corollary (see, e.g. \cite{Kerr:1963}):
After some algebra, we can show that
\begin{align}
&\left.\nabla_{a_1 \cdots a_r} I_{bcde}^{(1,1)}
\right|_{D{\boldsymbol K}=0} ~=~ 0
&&\Leftrightarrow
&&\left.\mathcal{L}_K \nabla_{a_1 \cdots a_r} R_{bcde}	
\right|_{D{\boldsymbol K}=0} ~=~ 0\:.
\end{align}
where $\mathcal{L}_K$ is the Lie derivative along a KV $K^a$.
This implies that if $Q$ is the scalar constructed out of
the Riemann curvature tensor and its derivatives,
then $\mathcal{L}_K Q$ must be zero.
So if the set of the $1$-form $\{dQ^{(1)}, \ldots, dQ^{(n)} \}$
are linearly independent, the $n$-form
\begin{align}
dQ^{(1)} \wedge \cdots \wedge dQ^{(n)}\:,
\end{align}
must also be zero for $n$-dimensional metrics and
is called a {\it curvature obstruction}.
If any of the possible obstructions is not vanishing, such a metric admits no KV.
We therefore expect that
further analysis of the explicit forms of
the integrability condition for $p>1$
will lead to similar obstructions for KTs.

A natural question to ask is whether we can formulate the existence condition or hopefully
the value of $r_0$ in Eq.\ \eqref{eq:r_0}.
Answering this question may be linked to the conjecture we made in Section \ref{sec:int_cond}.
In fact,
if the conjecture holds true, we obtain a criteria
\begin{align}
C~=~
\frac{\mathrm{rank}~I^{(p,p)}}{\mathrm{rank}~E^{(p)}}
~=~ \frac{n (n-1)}{(p+2) (p+1)} \:,
\label{eq:criteria}
\end{align}
where $n$ and $p$ are the dimension of
space(-time) $M$ and the order of KTs, respectively.
Here, $\mathrm{rank}~E^{(p)}$ agrees with the upper limit of
the BDTT formula \eqref{eq:dim_of_KS};
$\mathrm{rank}~I^{(p,p)}$ denotes
the number of linearly independent components of
the integrability condition \eqref{eq:int_pp}.
If $C \leq 1$, we definitely need the derivatives of $I^{(p,p)}$
to determine 
the dimension of the space of KTs
and thus $r_0 > 0$.
The equality is attained when $n=p+2$.
Namely, our conjecture serves to formulate
the lower bound on the value of $r_0$.

As shown in \ref{app:Yano},
our analysis based on Young symmetrizers
has effective applications to other types of overdetermined PDE systems.
The immediate examples include
the Killing--Yano equation, the $(p,q)$-type Killing spinor equations
and massless higher-spin field equations in $4$-dimensional spacetimes.
To analyze the conformal Killing(--Yano) equations,
we need not irreducible representations of GL$(n)$ but
those of SO$(n)$. So it will be necessary to incorporate the trace operation into our analysis.
Such modifications have not been pursued in this paper but will be considered in the future.

It would be worthwile to comment the significance of our results in
application to computer program. In recent years various softwares
implemented with a computer algebraic system, such as {\it Mathematica} and
{\it Maple}, have been developed. Each software prepares many packages available for solving individual problems in mathematics and physics, and we then find packages for solving the Killing equation for Killing vector fields as
well as Killing and Killing--Yano tensor fields.
However most of these packages do not solve the Killing equations efficiently, as they merely use a built-in PDE solver without the integrability conditions.
For fairness it should be mentioned that we found one package
(e.g. {\it KillingVectors} in {\it Maple})
which does use the integrability conditions, albeit only for Killing vector fields and not for Killing and Killing-Yano tensor fields.
Hence, in order to make such packages more efficient especially for Killing tensor fields, our
results in Section \ref{sec:int_cond} are significant to provide the formulas of the
integrability conditions to be implemented.

Finally, we close this paper with a comment on the properties of Young symmetrizers. In this paper, we have used many properties on Young symmetrizers.
For example, Raicu's formula \eqref{eq:raicu} has been used repeatedly
to a simplification of the product of two Young symmetrizers.
We remark that this formula can be applied only to
the case when one of the two Young tableau is contained in the other one.
Thus we needed other formulas \eqref{eq:interchange} and \eqref{eq:perm}
on the product of two Young symmetrizers
when we calculated the integrability conditions.
To obtain the conditions for $p > 3$, other new formulas would be required.

\section*{Acknowledgment}
We thank Claude M. Warnick and David Kubiznak for a stimulating discussion in the early stage of this work.
We also thank Maciej Dunajski, Konstantin Heil, Uwe Semmelmann 
and Sanefumi Moriyama for bringing relevant works to our attention.
We are grateful to Claudiu Raicu and Yasushi Homma for helping us with understanding the ABCs of Young symmetrizers.
We are also grateful to Vladimir S. Matveev for his useful comments.
This work was partially supported JSPS KAKENHI Grant Number
JP14J01237 (T. Houri) and JP16K05332 (Y. Yasui).

\appendix
\section{Young symmetrizer}\label{app:Young}
The aim of this appendix is to introduce the reader to 
some properties and formulas of Young symmetrizers
which play central roles in our calculations in 
\ref{app:prolong}.
See Ref.\ \cite{Fulton:2004} for more on
Young tableaux and the representation theory of symmetric groups.

\subsection{Basic properties}
We briefly present basic properties of Young symmetrizers defined by Eq.\ \eqref{def:Young}.
In our notation the Latin letters $(a,b,c, \ldots)$ are
identified as a naturally ordered set $(1,2,3,\ldots)$.
Therefore, for instance, the standard Young tableau
$Y_{{\tiny \begin{ytableau} 1 & 2\\ 3 & 4 \end{ytableau}}}\:$ is equated with
$Y_{{\tiny \begin{ytableau} a & b\\ c & d \end{ytableau}}}\:$
which is more suitable for tensor calculus.
We also order the subscripted Latin letters $(a_1, a_2, \ldots,b_1,b_2,\ldots)$
as $a_1 < a_2 <\cdots < b_1 <b_2<\cdots$.

In what follows, we shall denote the set of all standard Young tableaux with $k$ boxes
by $\mathcal{Y}_k$, e.g.
\begin{align*}
&\mathcal{Y}_2 ~=~ \bigl\{
{\tiny \begin{ytableau} a & b\end{ytableau}}~,
{\tiny \begin{ytableau} a \\ b\end{ytableau}}~
\bigr\}\:,
&&\mathcal{Y}_3 ~=~ \bigl\{
{\tiny \begin{ytableau} a & b & c\end{ytableau}}~,
{\tiny \begin{ytableau} a & b \\ c\end{ytableau}}~,
{\tiny \begin{ytableau} a & c \\ b\end{ytableau}}~,
{\tiny \begin{ytableau} a \\ b \\ c\end{ytableau}}~
\bigr\}\:.
\end{align*}

Young symmetrizers are endowed with the following three properties.
\begin{description}
	\item[Idempotence]
	\begin{align}
	&Y_{\Theta}^2 ~=~ Y_{\Theta}\:,
	&&\:^\forall \Theta \in \mathcal{Y}_k\:.
	\label{eq:idempotence}
	\end{align}
	\item[Orthogonality]
	\begin{align}
	&&Y_{\Theta}~Y_{\Phi} ~=~ \delta_{\Theta \Phi}~Y_{\Phi}\:,
	&&\:^\forall \Theta\:,\Phi \in \mathcal{Y}_k\:,
	&&(k=1,2,3,4).
	\label{eq:orthogonality}
	\end{align}
	\item[Completeness]
	\begin{align}
	&\sum_{\Theta \in \mathcal{Y}_k}~Y_{\Theta} ~=~ \mathrm{id}_k\:,
	&&(k=1,2,3,4).
	\label{eq:completeness}
	\end{align}
\end{description}
It is worth mentioning that the orthogonality \eqref{eq:orthogonality} holds for general $k$
if the shapes of the tableaux $\Theta$ and $\Phi$ are different.

For tensor calculus, 
the compleness relation \eqref{eq:completeness} plays an important role.
For instance, a decomposition of a tensor field of order $2$
into an irreducible representation of the general linear group GL$(2)$ can be done as follows.
\begin{align*}
T_{ab} ~=~ (
Y_{{\tiny \begin{ytableau} a & b\end{ytableau}}}
+ Y_{{\tiny \begin{ytableau} a \\ b\end{ytableau}}})T_{ab}
~=~T_{(ab)} + T_{[ab]}\:.
\end{align*}
For a tensor field of order $3$, a similar calculation reads
\begin{align*}
T_{abc} ~=~ (
Y_{{\tiny \begin{ytableau} a & b & c\end{ytableau}}}
+Y_{{\tiny \begin{ytableau} a & b \\ c\end{ytableau}}}
+Y_{{\tiny \begin{ytableau} a & c \\ b\end{ytableau}}}
+Y_{{\tiny \begin{ytableau} a \\ b \\ c\end{ytableau}}}
)T_{abc}
~=~ T_{(abc)} + \tfrac{4}{3}~\hat{A}_{ac} T_{(ab)c}
+ \tfrac{4}{3}~\hat{A}_{ab} T_{(a|b|c)}
+ T_{[abc]}
\:.
\end{align*}
But at $k=5$ the subsequent calculation reaches a deadlock
since the completeness relation \eqref{eq:completeness} no longer holds for $k\geq 5$.
The standard example of this is
\begin{align*}
&
Y_{{\tiny \begin{ytableau} a & b \\ c & d \\ e\end{ytableau}}}
~Y_{{\tiny \begin{ytableau} a & d \\ b & e \\ c\end{ytableau}}} ~=~0
&&\text{but}
&&
Y_{{\tiny \begin{ytableau} a & d \\ b & e \\ c\end{ytableau}}}
~Y_{{\tiny \begin{ytableau} a & b \\ c & d \\ e\end{ytableau}}} ~\neq~0\:.
\end{align*}

\subsection{Littlewood's correction}\label{sec:littlewood}
In practical use, the failure of the orthogonality and completeness of the Young symmetrizers for $k\geq 4$ is crucial. This failure is complemented by {\it Littlewood's correction} \cite{Littlewood:1950}.
We shall present it here. For other prescriptions,
see Refs. \cite{Stembridge:2011, Keppeler:2013}.

Before going into the details, we introduce the following two definitions:

\begin{definition}[row-word of a Young tableau]
	Let $\Theta \in \mathcal{Y}_k$ be a Young tableau.
	The row-word of $\Theta$, say $\mathrm{row}(\Theta)$,
	is defined as the row vector whose entries are those of $\Theta$
	read row-wise from top to bottom.
\end{definition}

For instance, suppose $\Theta = {\tiny \begin{ytableau} a & b \\ c & d \\ e\end{ytableau}}\:$.
Then the row-word of $\Theta$ reads $\mathrm{row} (\Theta) = (a, b, c, d, e)$.

\begin{definition}[row-order relation]
		Let $\Theta$ and $\Phi$ be two Young tableaux of the same shape.
		Denoting $\mathrm{row}(\Theta)_i$
		be the $i$-th component of $\mathrm{row}(\Theta)$,
		it is said that $\Theta$ precedes $\Phi$ and write $\Theta \prec \Phi$
		if $\mathrm{row} (\Theta)_i < \mathrm{row} (\Phi)_i$
		for the leftmost $i$ where $\mathrm{row} (\Theta)_i$ and $\mathrm{row} (\Phi)_i$ differ.
	\end{definition}
Using the row-order relation, we can order the Young tableaux of the same shape, e.g.
\begin{align*}
&{\tiny \begin{ytableau} a & b \\ c & d \\ e\end{ytableau}}
&&\prec
&&{\tiny \begin{ytableau} a & b \\ c & e \\ d\end{ytableau}}
&&\prec
&&{\tiny \begin{ytableau} a & c \\ b & d \\ e\end{ytableau}}
&&\prec
&&{\tiny \begin{ytableau} a & c \\ b & e \\ d\end{ytableau}}
&&\prec
&&{\tiny \begin{ytableau} a & d \\ b & e \\ c\end{ytableau}}.
\end{align*}
The following result is an easy consequence of the row-order relation:
Let $\{\Theta_1, \Theta_2, \Theta_3,\ldots\}$ be the set of all Young tableaux in $\mathcal{Y}_k$
with a particular shape.
Suppose this set be ordered as $\Theta_i \prec \Theta_j$ whenever $i<j$,
one can see by inspection that the one-sided orthogonality
\begin{align}
Y_{\Theta_i}~Y_{\Theta_j}~ = 0\:,
\label{eq:one-sided}
\end{align}
holds.

We are now able to state Littlewood's correction. The Young symmetrizer with Littlewood's correction, say $L_{\Theta_i}$,
corresponding the tableau $\Theta_i \in \{\Theta_1, \Theta_2, \Theta_3,\ldots\}$
is iteratively defined by
\begin{align}
L_{\Theta_i} ~\equiv~ Y_{\Theta_i}~
\Bigl(1-\sum_{j=1}^{i-1} L_{\Theta_j}\Bigr)\:,
\label{def:littlewood}
\end{align}
or the factorized form
\begin{align}
L_{\Theta_i} ~=~ Y_{\Theta_i} ~ 
\prod_{j=1}^{i-1}\Bigl(1-Y_{\Theta_{i-j}}\Bigr)\:.
\end{align}
As proven in Ref.\ \cite{Littlewood:1950},
the Young symmetrizers with correction \eqref{def:littlewood} recover the orthogonality,
\begin{align}
&&L_{\Theta}~L_{\Phi} ~=~ \delta_{\Theta \Phi}~L_{\Phi}\:,
&&\:^\forall \Theta\:,\Phi \in \mathcal{Y}_k\:,
\label{eq:orthogonality_L}
\end{align}
and the completeness,
\begin{align}
&\sum_{\Theta \in \mathcal{Y}_k}~L_{\Theta} ~=~ \mathrm{id}_k\:,
\label{eq:completeness_L}
\end{align}
for general $k$.

We note that all the corrections in \eqref{def:littlewood} vanish for $k \leq 4$, so that $L_{\Theta_i}$ are equivalent to $Y_{\Theta_i}$.
Even for $k > 4$, many corrections vanish. For example, for $k=5$, only two symmetrizers
\begin{align*}
&
L_{{\tiny \begin{ytableau} a & c & e \\ b & d\end{ytableau}}}
~=~Y_{{\tiny \begin{ytableau} a & c & e \\ b & d\end{ytableau}}}
~\Bigl( 1-
Y_{{\tiny \begin{ytableau} a & b & c \\ d & e\end{ytableau}}}
\Bigr)~,
&&
L_{{\tiny \begin{ytableau} a & d \\ b & e \\ c\end{ytableau}}}
~=~
Y_{{\tiny \begin{ytableau} a & d \\ b & e \\ c\end{ytableau}}}
~\Bigl( 1- 
Y_{{\tiny \begin{ytableau} a & b \\ c & d \\ e\end{ytableau}}} \Bigr)~,
\end{align*}
differ from their original counterparts. Thus it is useful for practical use to make it clear what kinds of the Young symmetrizes with
Littlewood's correction are equivalent to the original counterparts.
Since the tableau $
{{\tiny \begin{ytableau} a_1 &\none[...] &\none[...] &\none[...] &a_p\\
		b_1 & \none[...] & b_q \\
		c
		\end{ytableau}}}$
is row-ordered, it follows from the definition
\begin{align}
L_{{\tiny \begin{ytableau} a_1 &\none[...] &\none[...] &\none[...] &a_p\\
			b_1 & \none[...] & b_q \\
			c
			\end{ytableau}}}
~=~ Y_{{\tiny \begin{ytableau} a_1 &\none[...] &\none[...] &\none[...] &a_p\\
			b_1 & \none[...] & b_q \\
			c
			\end{ytableau}}}\:.
\label{eq:formula_101}
\end{align}
It is also shown that
\begin{align}
L_{{\tiny \begin{ytableau} a_1 &\none[...] &\none[...] &\none[...] &\none[...] &\none[...] &a_p & b_2\\
		b_1 & b_3& \none[...] & b_q & c\\
		\end{ytableau}}}
&=~
Y_{{\tiny \begin{ytableau} a_1 &\none[...] &\none[...] &\none[...] &\none[...] &\none[...] &a_p & b_2\\
		b_1 & b_3& \none[...] & b_q & c\\
		\end{ytableau}}}
\bigl( 1-
Y_{{\tiny \begin{ytableau} a_1 &\none[...] &\none[...] &\none[...] &\none[...] &\none[...] &a_p & b_1\\
		b_2 & b_3& \none[...] & b_q & c\\
		\end{ytableau}}}
\bigr)
~=~ Y_{{\tiny \begin{ytableau} a_1 &\none[...] &\none[...] &\none[...] &\none[...] &\none[...] &a_p & b_2\\
		b_1 & b_3& \none[...] & b_q & c\\
		\end{ytableau}}}\:,
\label{eq:formula_102}\\
L_{{\tiny \begin{ytableau} a_1 &\none[...] &\none[...] &\none[...]&\none[...] &\none[...] &\none[...] &a_p & b_3\\
		b_1 & b_2& b_4& \none[...] & b_q & c\\
		\end{ytableau}}}
&=~Y_{{\tiny \begin{ytableau} a_1 &\none[...] &\none[...] &\none[...]&\none[...] &\none[...] &\none[...] &a_p & b_3\\
		b_1 & b_2& b_4& \none[...] & b_q & c\\
		\end{ytableau}}}
\bigl( 1-
Y_{{\tiny \begin{ytableau} a_1 &\none[...] &\none[...] &\none[...] &\none[...] &\none[...] &a_p & b_1\\
		b_2 & b_3& \none[...] & b_q & c\\
		\end{ytableau}}}
\bigr)~
\bigl( 1-
Y_{{\tiny \begin{ytableau} a_1 &\none[...] &\none[...] &\none[...] &\none[...] &\none[...] &a_p & b_2\\
		b_1 & b_3& \none[...] & b_q & c\\
		\end{ytableau}}}
\bigr)
~=~Y_{{\tiny \begin{ytableau} a_1 &\none[...] &\none[...] &\none[...]&\none[...] &\none[...] &\none[...] &a_p & b_3\\
		b_1 & b_2& b_4& \none[...] & b_q & c\\
		\end{ytableau}}}\:, \label{eq:formula_103}
\end{align}
and so on, where we have only used the relations
$\hat{S}_{a_1 b_2}~\hat{A}_{a_1 b_2} = 0$ and
$\hat{S}_{a_2 b_3}~\hat{A}_{a_2 b_3} = 0$.
In general, the corrected symmetrizer
\begin{align}
&L_{{\tiny \begin{ytableau} a_1 &\none[...] &\none[...] &\none[...]&\none[...] &\none[...] &\none[...] &a_p & b_i\\
		b_1 &\none[...] & \Slash{b}_i & \none[...] &b_q & c\\
		\end{ytableau}}}\:,
&&\text{with}
&&p \geq q \geq 1\:,
&&q \geq i \geq 2\:,
\end{align}
is coincident with its original counterpart
by a trivial relation $\hat{S}_{a_{i-1} b_i}~\hat{A}_{a_{i-1}b_i} = 0$.


\subsection{Useful formulas}\label{app:formulae}
We collect some useful formulas
to perform calculations in 
\ref{app:prolong}.
	
We first glance at the result referred to as Schur's lemma
in the context of the representation theory of symmetric groups.
		Let $\Theta$ and $\Phi$ be Young tableaux with $k$ boxes.
		If $Y_\Theta$ and $Y_\Phi$ are orthogonal, that is
		$Y_\Theta ~ Y_\Phi = 0$, then
		\begin{align}
		Y_\Theta ~\sigma~ Y_\Phi = 0\:,
		\label{eq:schur}
		\end{align}
		holds true for an arbitrary permutation $\sigma$.

Second, we state Raicu's formula as follows.
	\footnote{
	To be precise, this result is an example of Raicu's theorem.
	A complete wording of Raicu's theorem can be found in Ref. \cite{Raicu:2013}}
		Let $\Theta \in \mathcal{Y}_k$ and $\Phi \in \mathcal{Y}_{k+1}$.
		Suppose that the unique entry in $\Phi$ outside $\Theta$ is
		located in the right edge of the tableau $\Phi$, then
		\begin{align}
		Y_{\Theta}~Y_{\Phi} ~=~ Y_{\Phi}\:,
		\label{eq:raicu}
		\end{align}
		holds.
	
	Next, we state an important result from the representation theory of symmetric groups.
		Let $\theta$ and $\phi$ be two Young diagrams with $k$ and $k+1$ boxes respectively.
		It is said that $\phi$ includes $\theta$ and write $\theta ~ \subset ~ \phi$
		if $\theta$ is a subdiagram of $\phi$.
		Let $\Theta$ and $\Phi$ be Young tableaux of shapes $\theta$ and $\phi$ respectively,
		then
		\begin{align}
		Y_{\Theta}~Y_{\Phi} ~=~ Y_{\Phi}~Y_{\Theta} ~=~ 0\:,
		\label{eq:pieri}
		\end{align}
		holds if $\phi$ does {\it not} include $\theta$.
		This result is called Pieri's formula.
		
		It should be noted that
		the first Bianchi identity, $R_{[abc]}\:^d = 0$,
		can be recaptured by Pieri's formula.
		We know that $R_{abcd}$ belongs to
		${\tiny \begin{ytableau} a & c\\ b & d\\
		\end{ytableau}}\:$,
		and hence the first Bianchi identity can be written in terms of
		Young symmetrizers as
		\begin{align*}
		Y_{\tiny \begin{ytableau} a \\ b\\ c
		\end{ytableau}}~
		Y_{\tiny \begin{ytableau} a & c\\ b & d\end{ytableau}}~=~0\:,
		\end{align*}
		which is clearly a type of Pieri's formula.
		Therefore, we can say that Pieri's formula
		is a generalization of the Bianchi identity.
		
At last, we state our findings
as pertains to the Young diagram of shape $(p+1, p+1)$:
\begin{align}
Y_{\tiny \begin{ytableau} a_1 & \none[...]& a_p \\
	b_1 & \none[...]& b_p
	\end{ytableau}}
~\Bigl[ 1+(-1)^{p-1}\prod_{i=1}^p (a_i, b_i)\Bigr] ~&=~0\:,
\label{eq:interchange}\\
Y_{\tiny \begin{ytableau} a_1 & \none[...]& a_p \\
	b_1 & \none[...]& b_p
	\end{ytableau}}
~\Bigl[ 1+\sum_{i=1}^{p-1} (a_1, b_i)\Bigr]~
Y_{\tiny \begin{ytableau} a_p \\
	b_p\end{ytableau}}~&=~0\:,
\label{eq:perm}
\end{align}
where $(a,b)$ denotes the permutation that swaps indices $a$ and $b$.

These formulas can be applicable to a simplification of the product of two Young symmetrizers
or of the operands of the rectangular Young symmetrizer.
We look at three examples before closing this appendix.
\begin{itemize}
\item[(i)] Consider the product
	$Y_{\tiny \begin{ytableau} a & b \\c & d
	\end{ytableau}}
	~Y_{\tiny \begin{ytableau} a & b \\c
		\end{ytableau}}\:$.
	It is simplified to
\begin{align}
Y_{\tiny \begin{ytableau} a & b \\c & d\end{ytableau}}
~Y_{\tiny \begin{ytableau} a & b \\c\end{ytableau}}
~=&~
Y_{\tiny \begin{ytableau} a & b \\c & d\end{ytableau}}
~Y_{\tiny \begin{ytableau} a & b \\c\end{ytableau}}
~\sum_{\Theta \in \mathcal{Y}_4}~Y_\Theta
~=~
Y_{\tiny \begin{ytableau} a & b \\c & d\end{ytableau}}
~Y_{\tiny \begin{ytableau} a & b \\c\end{ytableau}}
~Y_{\tiny \begin{ytableau} a & b \\c & d\end{ytableau}}
~=~
Y_{\tiny \begin{ytableau} a & b \\c & d\end{ytableau}}
~Y_{\tiny \begin{ytableau} a & b \\c & d\end{ytableau}}
~=~
Y_{\tiny \begin{ytableau} a & b \\c & d\end{ytableau}}\:,
\label{eq:demo1}
\end{align}
where the second and third equalities follow from Schur's lemma \eqref{eq:schur}
and Raicu's formula \eqref{eq:raicu} respectively.
\item[(ii)] Another example is the product
	$Y_{\tiny \begin{ytableau} a & b \\c \end{ytableau}}
	~\sum_{\Theta \in \mathcal{Y}_4} Y_\Theta$.
	A similar simplification can be done as
	\begin{align}
	Y_{\tiny \begin{ytableau} a & b \\c \end{ytableau}}
	~\sum_{\Theta \in \mathcal{Y}_4} Y_\Theta
	~=&~
	Y_{\tiny \begin{ytableau} a & b \\c \end{ytableau}}~
	\sum \Bigl(
	Y_{\tiny \begin{ytableau} \: & \: \\\: &\: \end{ytableau}}
	+Y_{\tiny \begin{ytableau} \: & \: \\\: \\\: \end{ytableau}}
	+Y_{\tiny \begin{ytableau} \: & \: &\: \\\: \end{ytableau}}
	\Bigr)
	~=~
	Y_{\tiny \begin{ytableau} a & b \\c \end{ytableau}}~
	\Bigl(
	Y_{\tiny \begin{ytableau} a & b \\c & d\end{ytableau}}
	+Y_{\tiny \begin{ytableau} a & b \\c \\ d\end{ytableau}}
	+Y_{\tiny \begin{ytableau} a & b &c \\ d\end{ytableau}}
	+Y_{\tiny \begin{ytableau} a & b &d \\ c\end{ytableau}}
	\Bigr)\:,
	\label{eq:demo2}
	\end{align}
	where in the first equality Pieri's formula \eqref{eq:pieri} have been used.
	\item[(iii)]
	A simplification of the operands of
	the rectangular Young symmetrizer goes as follows.
	\begin{align*}
	&Y_{\tiny \begin{ytableau} a & b &c & d\end{ytableau}}
	~\bigl[(\nabla_a R_{bcd}\:^m) K_m
	+(\nabla_c R_{dab}\:^m) K_m \bigr]
	~=~Y_{\tiny \begin{ytableau} a & b &c & d\end{ytableau}}
	~\bigl[ 2(\nabla_a R_{bcd}\:^m) K_m\bigr]\:,
	\end{align*}
	where the formula \eqref{eq:interchange} has been used.
	Another example is
	\begin{align*}
	&
	Y_{\tiny \begin{ytableau} a & b &c & g\\
		d & e & f & h\end{ytableau}}~\bigl[
	R_{ade}\:^m (\nabla_b R_{c fh}\:^n) K_{mn g}
	-R_{ade}\:^m (\nabla_b R_{fcg}\:^n) K_{mn h}
	\bigr]
	~=~	Y_{\tiny \begin{ytableau} a & b &c & g\\
		d & e & f & h\end{ytableau}}~\bigl[
	R_{ade}\:^m (\nabla_f R_{hbc}\:^n) K_{mn g}
	\bigr]\:,
	\end{align*}
	where we have applied the formula \eqref{eq:perm}.
\end{itemize}

\section{Prolongation procedure}\label{app:prolong}
In this appendix we shall demonstrate a prolongation procedure
for KTs of order $1$ and $2$.
Our calculation below can be extended to the higher order cases
that we have skipped here due to space considerations.
Throughout this appendix,
we repeatedly use the properties and formulas
of Young symmetrizers shown in 
\ref{app:Young}.

\subsection{Killing vector fields}
We start our investigation in the case of Killing vector $K^a$.
Calculating the derivative of $K_a$ can be done as
\begin{align*}
\nabla_b K_a ~=~
\bigl(
Y_{\tiny \begin{ytableau} a & b \end{ytableau}}
+Y_{\tiny \begin{ytableau} a \\ b \end{ytableau}}
\bigr)
\nabla_b K_a
~=~ Y_{\tiny \begin{ytableau} a \\ b \end{ytableau}}~
\nabla_b K_a
~\equiv~ K^{(1)}_{ba} \:,
\end{align*}
where we have inserted the completeness relation \eqref{eq:completeness}
in the first equality and
used the Killing equation in the second.

And calculating the derivative of $K^{(1)}_{ba}$ goes as
\begin{align*}
\nabla_c K^{(1)}_{ba} ~&=~
Y_{\tiny \begin{ytableau} a \\ b \end{ytableau}}
~\nabla_{cb} K_a
~=~
Y_{\tiny \begin{ytableau} a \\ b \end{ytableau}}~
\bigl(
Y_{\tiny \begin{ytableau} a & b & c\end{ytableau}}
+Y_{\tiny \begin{ytableau} a & b \\ c\end{ytableau}}
+Y_{\tiny \begin{ytableau} a & c \\ b\end{ytableau}}
+Y_{\tiny \begin{ytableau} a \\ b \\ c\end{ytableau}}
\bigr)
\nabla_{cb} K_a
~=~
Y_{\tiny \begin{ytableau} a \\ b \end{ytableau}}~
Y_{\tiny \begin{ytableau} a & c \\ b\end{ytableau}}~
\nabla_{cb} K_a \\
~&=~
Y_{\tiny \begin{ytableau} a \\ b \end{ytableau}}~
Y_{\tiny \begin{ytableau} a & c \\ b\end{ytableau}}~
\bigl(2\nabla_{[cb]} K_a + \nabla_{bc} K_a\bigr)
~=~
Y_{\tiny \begin{ytableau} a \\ b \end{ytableau}}~
Y_{\tiny \begin{ytableau} a & c \\ b\end{ytableau}}~
R_{cba} \: ^d K_d\:.
\end{align*}
In the forth equality,
no term has survived besides the third
because of the Killing equation and the first Bianchi identity.
The above calculations confirm the results \eqref{eq:prolong_10} and \eqref{eq:prolong_11}.

\subsection{Killing tensor fields of order $2$}
We shift our focus to the Killing tensor field $K^{ba}$.
Calculating the derivative of $K_{ba}$ gives
\begin{align*}
\nabla_c K_{ba}
&~=~
Y_{\tiny \begin{ytableau} a & b \end{ytableau}}~
\nabla_c K_{ba}
~=~
Y_{\tiny \begin{ytableau} a & b \end{ytableau}}~
\bigl(
Y_{\tiny \begin{ytableau} a & b & c\end{ytableau}}
+Y_{\tiny \begin{ytableau} a & b \\ c\end{ytableau}}
+Y_{\tiny \begin{ytableau} a & c \\ b\end{ytableau}}
+Y_{\tiny \begin{ytableau} a \\ b \\ c\end{ytableau}}
\bigr)
\nabla_c K_{ba} \\
&~=~
Y_{\tiny \begin{ytableau} a & b \end{ytableau}}~
Y_{\tiny \begin{ytableau} a & b \\ c\end{ytableau}}
\nabla_c K_{ba}
~=~
Y_{\tiny \begin{ytableau} a & b \end{ytableau}}~
K^{(1)}_{cba}\:,
\end{align*}
where the third equality follows from
the Killing equation and
a trivial relation $\hat{S}_{ab}~\hat{A}_{ab}=0$.

And calculating the derivative of $K^{(1)}_{cba}$ goes as
\begin{align}
\nabla_d K^{(1)}_{cba}
&~=~
Y_{\tiny \begin{ytableau} a & b \\ c\end{ytableau}}
~\nabla_{dc} K_{ba}
~=~
Y_{\tiny \begin{ytableau} a & b \\ c\end{ytableau}}~
\bigl(
Y_{\tiny \begin{ytableau} a & b \\ c \\ d\end{ytableau}}
+Y_{\tiny \begin{ytableau} a & b & d\\ c \end{ytableau}}
+Y_{\tiny \begin{ytableau} a & b \\ c & d\end{ytableau}}
\bigr)
\nabla_{dc} K_{ba} \notag\\
&~=~
Y_{\tiny \begin{ytableau} a & b \\ c\end{ytableau}}~
\bigl( \bigl[ Y_{\tiny \begin{ytableau} a & b \\ c \\ d\end{ytableau}}
+2Y_{\tiny \begin{ytableau} a & b \\ c & d\end{ytableau}}
\bigr]\nabla_{[dc]} K_{ba}
+K^{(2)}_{dcba}
\bigr)
~=~
Y_{\tiny \begin{ytableau} a & b \\ c\end{ytableau}}~
\bigl( \bigl[ Y_{\tiny \begin{ytableau} a & b \\ c \\ d\end{ytableau}}
+2Y_{\tiny \begin{ytableau} a & b \\ c & d\end{ytableau}}
\bigr]R_{dcb}\:^m K_{ma}
+K^{(2)}_{dcba}
\bigr)\:,
\label{eq:KS2_demo0}
\end{align}
where in the second equality
the result \eqref{eq:demo2} and the Killing equation have been used.
Expanding the two Young symmetrizers
in parentheses in Eq.\ \eqref{eq:KS2_demo0}, we obtain
\begin{align*}
\nabla_d K^{(1)}_{cba}
	~=&~ Y_{\tiny \begin{ytableau} a & b \\ c\end{ytableau}}~
	\Bigl[ K^{(2)}_{dcba}
	- \tfrac{5}{2} R_{dac}\:^mK_{mb}
	- 2 R_{dab}\:^mK_{mc}
	+\tfrac{1}{2} R_{acb}\:^mK_{md}
	\Bigr] \:.
\end{align*}

At last, calculating the derivative of $K^{(2)}_{dcba}$ can be done as
\begin{align*}
\nabla_e K^{(2)}_{dcba}
&~=~
Y_{\tiny \begin{ytableau} a & b \\ c & d\end{ytableau}}~
\nabla_{edc} K_{ba}
~=~
Y_{\tiny \begin{ytableau} a & b \\ c & d\end{ytableau}}~
\bigl(
L_{\tiny \begin{ytableau} a & b \\ c & d \\ e\end{ytableau}}
+L_{\tiny \begin{ytableau} a & b & d \\ c & e\end{ytableau}}
+L_{\tiny \begin{ytableau} a & b & e \\ c & d\end{ytableau}}
\bigr)
\nabla_{edc} K_{ba} \\
&~=~
Y_{\tiny \begin{ytableau} a & b \\ c & d\end{ytableau}}~
\bigl(
Y_{\tiny \begin{ytableau} a & b \\ c & d\\ e\end{ytableau}}
+Y_{\tiny \begin{ytableau} a & b & d \\ c & e\end{ytableau}}
(1 -Y_{\tiny \begin{ytableau} a & b & c \\ d & e\end{ytableau}} )
+
Y_{\tiny \begin{ytableau} a & b & e \\ c & d\end{ytableau}}
(1 - Y_{\tiny \begin{ytableau} a & b & d \\ c & e\end{ytableau}})
(1 - Y_{\tiny \begin{ytableau} a & b & c \\ d & e\end{ytableau}})
\bigr)
\nabla_{edc} K_{ba} \\
&~=~
Y_{\tiny \begin{ytableau} a & b \\ c & d\end{ytableau}}~
\bigl(
Y_{\tiny \begin{ytableau} a & b \\ c & d \\ e\end{ytableau}}
+Y_{\tiny \begin{ytableau} a & b & d \\ c & e\end{ytableau}}
+Y_{\tiny \begin{ytableau} a & b & e \\ c & d\end{ytableau}}
\bigr)
\nabla_{edc} K_{ba}\:.
\end{align*}
Now the number of boxes of the tableaux exceeds $4$,
so we have need to take Littlewood's correction \eqref{def:littlewood} into account.
However,
all these corrections are dropped by trivial relations
$\hat{S}_{ad}~\hat{A}_{ad} = 0$ and
$\hat{S}_{be}~\hat{A}_{be} = 0$.
Hence, our calculation can be pursued as
\begin{align}
\nabla_e K^{(2)}_{dcba}
~=~&
Y_{\tiny \begin{ytableau} a & b \\ c & d\end{ytableau}}~
\bigl(
Y_{\tiny \begin{ytableau} a & b \\ c & d \\ e\end{ytableau}}
+Y_{\tiny \begin{ytableau} a & b & d \\ c & e\end{ytableau}}
+Y_{\tiny \begin{ytableau} a & b & e \\ c & d\end{ytableau}}
\bigr)
\nabla_{edc} K_{ba} \notag\\
~=~&
Y_{\tiny \begin{ytableau} a & b \\ c & d\end{ytableau}}~
\Bigl\{
Y_{\tiny \begin{ytableau} a & b \\ c & d \\ e\end{ytableau}}
\nabla_{edc} K_{ba}
+Y_{\tiny \begin{ytableau} a & b & d \\ c & e\end{ytableau}}
(2 \nabla_{e[dc]} K_{ba})
+Y_{\tiny \begin{ytableau} a & b & e \\ c & d\end{ytableau}}
(2 \nabla_{[ed]c} K_{ba} + 2 \nabla_{d[ec]} K_{ba})
\Bigr\} \notag\\
~=~&
Y_{\tiny \begin{ytableau} a & b \\ c & d\end{ytableau}}~
\biggl\{
Y_{\tiny \begin{ytableau} a & b \\ c & d \\ e\end{ytableau}}\nabla_{edc} K_{ba}
+Y_{\tiny \begin{ytableau} a & b & d\\ c & e\end{ytableau}}
\Bigl( 2(\nabla_e R_{dcb}\:^m) K_{ma}
+R_{dcb}\:^m K^{(1)}_{mea}
-2R_{dcb}\:^m K^{(1)}_{mae}
\Bigr) \notag \\
&
+Y_{\tiny \begin{ytableau} a & b & e\\ c & d\end{ytableau}}
\Bigl(R_{edc}\:^m K^{(1)}_{mba}
+ 2 R_{edb}\:^m K^{(1)}_{mca}
-4 R_{edb}\:^m K^{(1)}_{mac}
+2 (\nabla_d R_{ecb}\:^m) K_{ma}\Bigr)
\biggr\}\:.
\label{eq:KS2_demo1}
\end{align}
Regarding the first term, a straightforward calculation gives
\begin{align}
Y_{\tiny \begin{ytableau} a & b \\ c & d \\ e\end{ytableau}}
\nabla_{edc} K_{ba}
&~=~\tfrac{1}{6} \bigl(
4 R_{ecd}\:^m K^{(1)}_{mab}
-9 R_{acd}\:^m K^{(1)}_{meb}
-9 R_{eac}\:^m K^{(1)}_{mdb}
+5 R_{acd}\:^m K^{(1)}_{mbe}
+5 R_{eca}\:^m K^{(1)}_{mdb} \notag\\
&+2 (\nabla_c R_{eda}\:^m) K_{mb}
+2 (\nabla_c R_{dab}\:^m) K_{me}
-2(\nabla_c R_{eab}\:^m) K_{md}
\bigr)\:.
\label{eq:KS2_demo2}
\end{align}
Combining the results of \eqref{eq:KS2_demo1} and \eqref{eq:KS2_demo2}
leads to the conclusion that
\begin{align*}
\nabla_e K^{(2)}_{dcba}
~=~&
Y_{\tiny \begin{ytableau} a & b \\ c & d\end{ytableau}}~
\Bigl( -\tfrac{4}{3} (\nabla_a R_{bcd}\:^m) K_{me}
-\tfrac{2}{3} (\nabla_e R_{cab}\:^m) K_{md}
-\tfrac{8}{3} (\nabla_a R_{bde}\:^m) K_{mc} \notag\\
&-12 R_{eac}\:^m K_{mdb}^{(1)}
-4 R_{eab}\:^m K_{mcd}^{(1)}
-\tfrac{2}{3} R_{cab}\:^m K_{mde}^{(1)}
+\tfrac{7}{3} R_{cab}\:^m K_{med}^{(1)}
\Bigr)\:,
\end{align*}
where we have expanded the three Young symmetrizers
in parentheses in Eq.\ \eqref{eq:KS2_demo1} explicitly.

\section{Derivation of the integrability conditions}\label{app:d_int}
In this appendix we shall verify the integrability conditions \eqref{eq:int_pp}.
We begin with the integrability conditions of the $p$th prolonged equation.
Evaluating the expression \eqref{eq:tensor_i} at $q=p$, one finds that
\begin{align}
I^{(p,p)}_{a_1 \cdots a_p b_1 \cdots b_p cd}
~=~&
Y_{\tiny \begin{ytableau} c \\ d\end{ytableau}}~
Y_{\tiny \begin{ytableau} a_1 & \none[...] & a_p \\
b_1 & \none[...] & b_p\end{ytableau}}~
\biggl[
Y_{\tiny \begin{ytableau} a_1 & \none[...] & a_p \\
	b_1 & \none[...] & b_p\\ c\end{ytableau}}
\nabla_{dc b_p \cdots b_1} K_{a_p \cdots a_1}
+Y_{{\tiny \begin{ytableau}
		a_1 &\none[...] &a_p & c\\
		b_1 & \none[...] & b_p
		\end{ytableau}}}~
\nabla_{d c b_p\cdots b_1} K_{a_p \cdots a_1} \notag\\
&
+\sum^{p}_{i=2}
Y_{{\tiny \begin{ytableau}
		a_1 &\none[...] &\none[...] &\none[...] &\none[...] &a_p&b_i\\
		b_1 & \none[...] &\Slash{b}_i &\none[...]& b_p & c
		\end{ytableau}}}~
\nabla_{d c b_p\cdots b_1} K_{a_p \cdots a_1}
-\nabla_{[dc]} K^{(p)}_{b_p \cdots b_1a_p \cdots a_1}
\biggr]
\:.
\label{eq:int_proof0}
\end{align}
The last term in Eq.\ \eqref{eq:int_proof0} can be treated as
\begin{align*}
Y_{\tiny \begin{ytableau} c \\ d\end{ytableau}}~
Y_{\tiny \begin{ytableau} a_1 & \none[...] & a_p \\
	b_1 & \none[...] & b_p\end{ytableau}}~
\nabla_{[dc]} K^{(p)}_{b_p \cdots b_1 a_p \cdots a_1}
~=~&
Y_{\tiny \begin{ytableau} c \\ d\end{ytableau}}~
Y_{\tiny \begin{ytableau} a_1 & \none[...] & a_p \\
	b_1 & \none[...] & b_p\end{ytableau}}~
\sum_{\Theta \in \mathcal{Y}_{2p+1}} L_{\Theta}
~
\nabla_{[dc]} K^{(p)}_{b_p \cdots b_1 a_p \cdots a_1}\\
~=~&
Y_{\tiny \begin{ytableau} c \\ d\end{ytableau}}~
Y_{\tiny \begin{ytableau} a_1 & \none[...] & a_p \\
	b_1 & \none[...] & b_p\end{ytableau}}~
\Bigl(
Y_{\tiny \begin{ytableau} a_1 & \none[...] & a_p \\
	b_1 & \none[...] & b_p \\c\end{ytableau}}
+Y_{\tiny \begin{ytableau} a_1 & \none[...] & a_p & c\\
	b_1 & \none[...] & b_p \\c\end{ytableau}}
\Bigr)
\nabla_{[dc]} K^{(p)}_{b_p \cdots b_1 a_p \cdots a_1}\:,
\end{align*}
where Pieri's formula \eqref{eq:pieri} is used and
Littlewood's correction \eqref{def:littlewood} is dropped by
a relation $\hat{S}_{a_p c}~\hat{A}_{a_p c}~=~0$.
Hence, Eq.\ \eqref{eq:int_proof0} can be rewritten as
\begin{align}
I^{(p,p)}_{a_1 \cdots a_p b_1 \cdots b_p cd}
~=~&
Y_{\tiny \begin{ytableau} c \\ d\end{ytableau}}~
Y_{\tiny \begin{ytableau} a_1 & \none[...] & a_p \\
	b_1 & \none[...] & b_p\end{ytableau}}~
\biggl[
\Bigl(
Y_{\tiny \begin{ytableau} a_1 & \none[...] & a_p \\
	b_1 & \none[...] & b_p\\ c \end{ytableau}}
+Y_{\tiny \begin{ytableau} a_1 & \none[...] & a_p & c\\
	b_1 & \none[...] & b_p\end{ytableau}}
\Bigr)
\Bigl(
\nabla_{dc b_p \cdots b_1} K_{a_p \cdots a_1}
-\nabla_{[d c]} K^{(p)}_{b_{p} \cdots b_1 a_p\cdots a_1}
\Bigr)\notag\\
&~+\sum^{p}_{i=2}
Y_{{\tiny \begin{ytableau}
		a_1 &\none[...] &\none[...] &\none[...] &\none[...] &a_p&b_i\\
		b_1 & \none[...] &\Slash{b}_i &\none[...]& b_p & c
		\end{ytableau}}}~
\nabla_{dc b_p \cdots b_1} K_{a_p \cdots a_1}
\biggr]
\:.
\label{eq:int_proof1}
\end{align}

Suppose now that the conjecture in Section \ref{sec:int_cond} holds true.
We then ignore the first symmetrizer $Y_{\tiny \begin{ytableau} a_1 & \none[...] & a_p \\
	b_1 & \none[...] & b_p\\ c \end{ytableau}}$
since it does not induce the representations 
belonging to $(p+1, p+1)$.
The products of Young symmetrizers in Eq.\ \eqref{eq:int_proof1} can be simplified to
\begin{align}
Y_{\tiny \begin{ytableau} a_1 & \none[...] & a_p \\
	b_1 & \none[...] & b_p\end{ytableau}}~
Y_{\tiny \begin{ytableau} a_1 & \none[...] & a_p & c\\
	b_1 & \none[...] & b_p\end{ytableau}}
~=~&
Y_{\tiny \begin{ytableau} a_1 & \none[...] & a_p & c\\
	b_1 & \none[...] & b_p\end{ytableau}}~\:, 
\label{eq:int_proof_product0}\\
Y_{\tiny \begin{ytableau} a_1 & \none[...] & a_p \\
	b_1 & \none[...] & b_p\end{ytableau}}~
Y_{{\tiny \begin{ytableau}
		a_1 &\none[...] &\none[...] &\none[...] &\none[...] &a_p&b_i\\
		b_1 & \none[...] &\Slash{b}_i &\none[...]& b_p & c
		\end{ytableau}}}
~=~&\tfrac{1}{2}
Y_{{\tiny \begin{ytableau}
		a_1 &\none[...] &a_p&c\\
		b_1 & \none[...] & b_p
		\end{ytableau}}}
(c, b_p) \prod_{j=1}^{p-i}(b_{p+1-j}, b_{p-j})\:.
\label{eq:int_proof_product1}
\end{align}
The first result \eqref{eq:int_proof_product0} follows immediately from
Raicu's formula \eqref{eq:raicu}.
The second result \eqref{eq:int_proof_product1} can be confirmed by a direct calculation.
By using the relations \eqref{eq:int_proof_product0} and \eqref{eq:int_proof_product1},
the equation \eqref{eq:int_proof1} can be rewritten as
\begin{align}
&I^{(p,p)}_{a_1 \cdots a_p b_1 \cdots b_p cd}
~=~
Y_{{\tiny \begin{ytableau}c\\d
\end{ytableau}}}~
Y_{{\tiny \begin{ytableau}
		a_1 &\none[...] &a_p&c\\
		b_1 & \none[...] & b_p
		\end{ytableau}}}~
\mathrm{id}_{2p+2}\biggl[
\nabla_{dc b_p \cdots b_1} K_{a_p \cdots a_1}
+
\frac{1}{2}\nabla_{d b_p c b_{p-1}\cdots b_1} K_{a_p \cdots a_1} \notag \\
&+
\frac{1}{2}\nabla_{d b_p b_{p-1} c b_{p-2}\cdots b_1} K_{a_p \cdots a_1}
+\cdots +
\frac{1}{2}\nabla_{d b_p b_{p-1} b_{p-2} \cdots c b_1} K_{a_p \cdots a_1}
-\nabla_{[dc]} K^{(p)}_{b_p \cdots b_1 a_p \cdots a_1}
\biggr] \:.
\label{eq:int_proof2}
\end{align}
Expanding $\mathrm{id}_{2p+2}$
and the antisymmetrizations of
the operands yields the result \eqref{eq:int_pp}.

\section{Killing-Yano equation}\label{app:Yano}
Our analysis based on Young symmetrizers
has effective applications to other types of overdetermined PDE systems.
In this appendix we discuss the Killing-Yano equation
\begin{align}
\nabla_{(b} F_{a_1) \cdots a_p} ~=~ 0\:,
\label{eq:Killing-Yano}
\end{align}
where $F_{a_1\cdots a_p}=F_{[a_1\cdots a_p]}$ is a Killing-Yano tensor field (KY).
If we have a KY, then we can obtain a KT of order $2$ as
\begin{align}
K_{ab} ~\equiv~ F_{a c_1 \cdots c_{p-1}}~F_b{}^{c_1 \cdots c_{p-1}}\:,
\end{align}
but the converse is not generally true.
While the prolonged equations of the Killing-Yano equation
and their integrability conditions have been known in \cite{Houri:2015,Batista:2015a,Batista:2015b},
we revisit the results by using Young symmetrizers.

Let $F_{ab}$ be a KY and consider its derivatives.
Since $\nabla_c F_{ba}$ is a type $(0,3)$ tensor field,
its decomposition to the irreducible representations reads
\begin{align}
\nabla_c F_{ba} ~=~
Y_{{\tiny \begin{ytableau}a\\b
		\end{ytableau}}}~
\mathrm{id}_3~\nabla_c F_{ba}
~=~ Y_{{\tiny \begin{ytableau}a\\b
		\end{ytableau}}}~
\Bigl(
Y_{{\tiny \begin{ytableau}a\\b\\c
		\end{ytableau}}}
+Y_{{\tiny \begin{ytableau}a&b\\c
		\end{ytableau}}}
+Y_{{\tiny \begin{ytableau}a&c\\b
		\end{ytableau}}}
\Bigr)
~\nabla_c F_{ba}
~=~
Y_{{\tiny \begin{ytableau}a\\b\\c
		\end{ytableau}}}
~\nabla_c F_{ba}
~\equiv~F_{cba}^{(1)}\:,
\end{align}
where we have used Pieri's formula \eqref{eq:pieri}
and the Killing-Yano equation \eqref{eq:Killing-Yano}.
We next consider $\nabla_d F_{cba}^{(1)}$
as the above result is not yet closed.
Its decomposition to the irreducible representations reads
\begin{align}
\nabla_d F_{cba}^{(1)}&~=~
Y_{{\tiny \begin{ytableau}a\\b\\c
		\end{ytableau}}}
~\mathrm{id}_4~\nabla_{dc} F_{ba}
~=~
Y_{{\tiny \begin{ytableau}a\\b\\c
		\end{ytableau}}}~
\Bigl(
Y_{{\tiny \begin{ytableau}a\\b\\c\\d
		\end{ytableau}}}
+Y_{{\tiny \begin{ytableau}a&b\\c\\d
		\end{ytableau}}}
+Y_{{\tiny \begin{ytableau}a&c\\b\\d
		\end{ytableau}}}
+Y_{{\tiny \begin{ytableau}a&d\\b\\c
		\end{ytableau}}}
\Bigr)
~\nabla_{dc} F_{ba}
~=~Y_{{\tiny \begin{ytableau}a&d\\b\\c
		\end{ytableau}}}~\nabla_{dc} F_{ba} \notag \\
&~=~Y_{{\tiny \begin{ytableau}a&d\\b\\c
		\end{ytableau}}}~
\Bigl(
2\nabla_{[dc]} F_{ba} + \nabla_{cd} F_{ba}
\Bigr)
~=~2 Y_{{\tiny \begin{ytableau}a&d\\b\\c
		\end{ytableau}}}~Y_{{\tiny \begin{ytableau}a\\b
		\end{ytableau}}}~
R_{dcb}{}^mF_{ma}\:,
\end{align}
which is now closed.
This implies that we are at the completion of the procedure of prolongation.

A similar calculation,
taking into account Littlewood’s corrections, yields the conclusion
that the Killing-Yano equation \eqref{eq:Killing-Yano} is equivalent to
the prolonged equations
\ytableausetup{boxsize=6pt}
\begin{align}
\nabla_b F_{a_p \cdots a_1} &~=~ F^{(1)}_{ba_p \cdots a_1}\:,
\label{eq:KY_prolong1}
\\
\nabla_c F^{(1)}_{ba_p \cdots a_1} &~=~ p
Y_{{\tiny \begin{ytableau} a_1 &c \\
	\none[\rotatebox{90}{\hspace{-1pt}...}]\\
	a_p \\
	b
	\end{ytableau}}}~
Y_{{\tiny \begin{ytableau} a_1 \\
	\none[\rotatebox{90}{\hspace{-1pt}...}]\\
		a_p
		\end{ytableau}}}~R_{cba_p}{}^m F_{m a_{p-1}\cdots a_1}\:,
\label{eq:KY_prolong2}
\end{align}
where
\begin{align}
F^{(1)}_{ba_p \cdots a_1} ~\equiv~
Y_{{\tiny \begin{ytableau} a_1 \\
	\none[\rotatebox{90}{\hspace{-1pt}...}]\\
		a_p\\
		b
		\end{ytableau}}}~ \nabla_b F_{a_p \cdots a_1}\:.
\end{align}

	After a calculation analogous to that in \ref{app:d_int},
	we obtain the integrability condition for Eq.\ \eqref{eq:KY_prolong1}
\begin{align}
&Y_{{\tiny \begin{ytableau} a_1 &b \\
		\none[\rotatebox{90}{\hspace{-1pt}...}] & c\\
		\none[\rotatebox{90}{\hspace{-1pt}...}]\\
		a_p
		\end{ytableau}}}~
	\left[
	R^m{}_{ca_1\cdots}{} F_{\cdots a_p b m}
	\right]~=~0\:,
&&\text{for}
&&p>1\:.
	\label{eq:KY_int-cond}
\end{align}
It can be confirmed that 
the integrability condition for Eq.\ \eqref{eq:KY_prolong2}
is involved in the derivative of Eq.\ \eqref{eq:KY_int-cond}.
Therefore, Eq.\ \eqref{eq:KY_int-cond} and its derivatives are
enough to discuss the integrability condition
of the Killing--Yano equation.
Once again, we face a situation similar to the one just discussed in Section \ref{sec:int_cond}.
Namely, there is only a representation in Eq.\ \eqref{eq:KY_int-cond},
even though the possible representations of Eq.\ \eqref{eq:KY_int-cond} are three
\begin{align*}
&{\tiny \begin{ytableau} \quad\\ \none[\rotatebox{90}{\hspace{-1pt}...}]
	\\ \none[\rotatebox{90}{\hspace{-1pt}...}]\\ \quad \\ \quad \\ \quad\end{ytableau}}~,
&&{\tiny \begin{ytableau} \quad & \quad\\
\none[\rotatebox{90}{\hspace{-1pt}...}] \\
\none[\rotatebox{90}{\hspace{-1pt}...}]\\
\quad \\ \quad\end{ytableau}}~,
&&{\tiny \begin{ytableau} \quad &\quad \\
	\none[\rotatebox{90}{\hspace{-1pt}...}] & \quad\\
	\none[\rotatebox{90}{\hspace{-1pt}...}]\\
	\quad
	\end{ytableau}}~.
\end{align*}


\begin{thebibliography}{99}

\bibitem{Craig:2008}
W. Craig,
Hamiltonian Dynamical Systems and Applications,
Springer 2008.

\bibitem{Cariglia:2014review}
M. Cariglia,
Rev. Mod. Phys. {\bf 86}, 1283 (2014).


\bibitem{Stackel:1895}
P. St\"ackel,
C. R. Acd. Sci. Paris Ser. IV, {\bf 121} 489 (1895).

\bibitem{Benenti:1975}
S. Benenti,
Rend. Sem. Mat. Univ. Pol. Torino, 34 (1975-6), 431;
Rep. Math. Phys. {\bf 12}, 311-316 (1977).

\bibitem{Benenti:1980}
S. Benenti, M. Francaviglia,
Gen. Rel. Grav. {\bf 10}, 79-92 (1979);
The theory of separability of the Hamilton-Jacobi equation and its applications to General Relativity, IN:
Gen. Rel. Grav. vol. I, New York: Plenum Press (1980).

\bibitem{Carter:1968}
B. Carter, Commun. Math. Phys. {\bf 10}, 280-310 (1968).

\bibitem{Walker:1970}
M. Walker and R. Penrose,
Commun. Math. Phys. {\bf 18}, 265-274 (1970).


\bibitem{Davis:2005}
P. Davis, H. K. Kunduri, J. Lucietti,
Phys. Lett. {\bf B628}, 275-280 (2005).

\bibitem{Kubiznak:2007}
D. Kubiznak, V. P. Frolov,
Class. Quant. Grav. {\bf 24}, F1-F6 (2007).

\bibitem{Hioki:2008}
K. Hioki, U. Miyamoto,
Phys. Rev. {\bf D78}, 044007 (2008).

\bibitem{Houri:2010}
T. Houri, D. Kubiznak, C. M. Warnick, Y. Yasui,
JHEP {\bf 1007}, 055 (2010).

\bibitem{Chow:2010}
D. D. K. Chow,
Class. Quant. Grav. {\bf 27}, 205009 (2010).

\bibitem{Kruglikov:2012}
B. S. Kruglikov, V. S. Matveev,
Phys. Rev. {\bf D85}, 12, 124057 (2012).

\bibitem{Vollmer:2016}
A. Vollmer,
arXiv:1602.08968 [math.DG].

\bibitem{Chow:2016}
D. D. K. Chow,
arXiv:1608.05052 [hep-th].

\bibitem{Gibbons:2011}
G. W. Gibbons, T. Houri, D. Kubiznak, C. M. Warnick,
Phys. Lett. {\bf B700}, 68-74 (2011).

\bibitem{Galajinsky:2012}
A.\ Galajinsky,
Phys. Rev. {\bf D85}, 085002 (2012).

\bibitem{Cariglia:2014}
M. Cariglia, G. W.Gibbons, J.-W. van Holten, P. A. Horvathy, P. Kosinski, P.-M. Zhang,
Class. Quantum Grav. {\bf 31}, 125001 (2014).

\bibitem{Bryant:1991}
R. L. Bryant, S. S. Chern, R. B. Gardner, H. L. Goldschmidt, P. A. Griffiths,
Exterior Differential Systems,
Springer-Verlag (1991).

\bibitem{Eastwood:2008}
M. Eastwood, W. Miller, Jr.,
Symmetries and Overdetermined Systems of Partial Differential Equations,
Springer 2008.

\bibitem{Dunajski:2010}
Maciej Dunajski,
Solitions, Instantons and Twistors,
Oxford University Press (2010).

\bibitem{Wald:1984}
R. M. Wald,
General Relativity,
The University of Chicago Press, 1984.

\bibitem{Branson:2007}
T. Branson, A. Cap, M. Eastwood, R. Gover,
Int. J. Math. {\bf 17}, 641-664 (2007).

\bibitem{Michel:2014}
J.-P. Michel, P. Somberg and J. \v{S}ilhan,
arXiv:1403.7226.


\bibitem{Barbance:1973}
de-M. C. Barbance,
G. R. Acad. Sc. Paris, {\bf 276}, S\'erie A, 387-389 (1973).

\bibitem{Delong:1982}
R. P. Delong,
PhD Thesis, University of Minnesota (1982).

\bibitem{Takeuchi:1983}
M. Takeuchi,
Tsukuba J. Math. {\bf 7}, 233-255 (1983).

\bibitem{Thompson:1986}
G. Thompson,
J. Math. Phys. {\bf 27}, 2693-2699 (1986).


\bibitem{Veblen:1923}
O. Veblen and T. Y. Thomas,
The Geometry of Paths,
Trans. American Math. Soc. {\bf 25}, 551-608 (1923).

\bibitem{Hauser:1974}
I. Hauser and R. J. Malhiot,
J. Math. Phys. {\bf 15}, 816 (1974).

\bibitem{Hauser:1975i}
I. Hauser and R. J. Malhiot,
J. Math. Phys. {\bf 16}, 150 (1975).

\bibitem{Hauser:1975ii}
I. Hauser and R. J. Malhiot,
J. Math. Phys. {\bf 16}, 1625 (1975).

\bibitem{Wolf:1998}
T. Wolf,
Comp. Phys. Comm. {\bf 115}, 316 - 329 (1998).


\bibitem{Semmelmann:2002}
U. Semmelmann,
math/0206117.

\bibitem{Batista:2014}
C. Batista,
Class. Quantum Grav. {\bf 31}, 165019 (2014).

\bibitem{Houri:2015}
T. Houri, Y. Yasui,
Class. Quant. Grav. {\bf 32}, 055002 (2015).

\bibitem{Batista:2015a}
C. Batista,
Phys. Rev. {\bf D91}, 024013 (2015).

\bibitem{Batista:2015b}
C. Batista,
Phys. Rev. {\bf D91}, 084036 (2015).

\bibitem{Houri:2016}
T. Houri, Y. Morisawa, K. Tomoda,
J. Math. Phys. {\bf 57}, 022501 (2016).


\bibitem{Kruglikov:2016}
B. S. Kruglikov, V. S. Matveev,
Nonlinearity {\bf 29} (2016) 1755-1768.

\bibitem{Bryant:2008}
R.L. Bryant, M. Dunajski, M. Eastwood,
J. Diff. Geom. {\bf 83} (2009) no.3, 465–500.

\bibitem{Kerr:1963}
R. P. Kerr,
J. Math. Mech. {\bf 12}, 33 (1963).



\bibitem{Fulton:2004}
W. Fulton, J. Harris,
Representation Theory,
Springer 2004.

\bibitem{Littlewood:1950}
D. Littlewood,
The Theory of Group Characters and Matrix Representations of Groups. Clarendon,
UK: Oxford Univ. Press. 1950.

\bibitem{Stembridge:2011}
J. R. Stembridge,
Adv. Appl. Math. {\bf 46}, 576-582 (2011).

\bibitem{Keppeler:2013}
S. Keppeler, M. Sjodahl,
J. Math. Phys. {\bf 55}, 021702 (2014).

\bibitem{Raicu:2013}
C. Raicu,
J. Algebraic Combin. {\bf 39}, no. 2:247-270 (2014).

\end{thebibliography}
\end{document}